\documentclass[notitlepage, amsmath,amssymb,longbibliography,superscriptaddress,floatfix]{revtex4-1}

\usepackage{graphicx}
\usepackage{dcolumn}
\usepackage{color}
\usepackage{bm}
\usepackage{tabularx}
\usepackage{float}
\usepackage{afterpage}
\usepackage{rotating}

\usepackage[hidelinks]{hyperref}
\hypersetup{
    colorlinks,
    citecolor=blue,
    filecolor=blue,
    linkcolor=blue,
    urlcolor=blue
}

\newcommand\varpm{\mathbin{\vcenter{\hbox{%
  \oalign{\hfil$\scriptstyle+$\hfil\cr
          \noalign{\kern-.5ex}
          $\scriptscriptstyle({-})$\cr}%
}}}}

\newcommand\varmp{\mathbin{\vcenter{\hbox{%
  \oalign{\hfil$\scriptstyle-$\hfil\cr
          \noalign{\kern-.5ex}
          $\scriptscriptstyle({+})$\cr}%
}}}}

\newcolumntype{L}[1]{>{\raggedright\arraybackslash}p{#1}}
\newcolumntype{C}[1]{>{\centering\arraybackslash}p{#1}}
\newcolumntype{R}[1]{>{\raggedleft\arraybackslash}p{#1}}

\DeclareMathOperator{\sgn}{sgn}

\DeclareMathOperator*{\pr}{\parallel}
\DeclareMathOperator*{\arccosh}{arccosh}

\begin{document}

\title{Theory of universal differential conductance of magnetic Weyl type -II junctions}

\author{M. Maiti}
 \email{maiti@theor.jinr.ru}
\affiliation {Bogoliubov Laboratory of Theoretical Physics, Joint Institute for Nuclear Research, 141980 Dubna, Moscow Region, Russia} 
\author{J. Smotlacha}
\affiliation {Bogoliubov Laboratory of Theoretical Physics, Joint Institute for Nuclear Research, 141980 Dubna, Moscow Region, Russia} 
\affiliation {Faculty of Nuclear Sciences and Physical Engineering, Czech Technical University, Brehova 7, 110 00 Prague, Czech Republic}

\date{\today}

\begin{abstract}

We study the transport properties of junctions of normal and superconducting Weyl semi-metal with tilted dispersion,
in the presence of magnetization induced by magnetic strips. The sub gap tunnelling conductance shows
robust signatures in the presence of different orientation and strength of magnetization
of the magnetic strips. We obtain the analytical results for the normal-magnetic-superconducting
junction in the thin barrier limit and demonstrate that these results have
no analogues to their conventional counterparts and junctions with Dirac electrons in two-dimensions.
We discuss possible experimental setups to test our theoretical predictions.
\end{abstract}


\maketitle

\section{Introduction}

Weyl semimetals (WSM) is the latest addition to the list of materials which have attracted wide interests from
theoretical and experimental research alike due to the unique possibility of realization of elusive
and sought after relativistic fermions in table-top condensed matter systems. Weyl semimetals
are further classified into two groups in terms of Lorentz symmetry; a type-I which respects and the
type -II which breaks the Lorentz invariance. In WSM type-II semimetals,
a tilted band structure with open constant energy surfaces and symmetry protected pairs of Weyl nodes appears
at the contact of co-existing electron and holes pockets
\cite{Vishwanath2011, *Soluyanov2015, *Armitage2018}.
Material realizations of type-II Weyl material has been possible in layered transition
metal dichalcogenides which are strong spin-orbit coupled materials. Surface Fermi arcs which are
characteristic signatures of the Weyl dispersion has been observed in
$\rm LaAlGe$ \cite{Xue1603266}, $MoTe_2$ \cite{Deng2016, *Huang2016, *Jiang2017, *Sakono2017},
$WTe_2$ \cite{Kaminski2016}, $WP_2$ \cite{Mesot2019}.
Violation of Lorentz symmetry in Weyl semimetals gives rise to distinguished
transport features. Heterostructures of normal and superconducting Weyl II semimetals
are shown to exhibit double Andreev reflections \cite{Sun2017}, double electron cotunnelling \cite{Li2018},
signatures of non-local transport characterized by crossed Andreev reflections\cite{Li2019}.
These properties are in contrast to the WSM type-I junctions
\cite{Chen2013, *Madsen2017, *Khanna2016, *Khanna2017, *Mukherjee2017, *Zhang2018, *Sinha2019}, or
junctions hosting Dirac fermions \cite{Beenakker2006, *Bhattacharjee2006, *Maiti2007}.

External electric and magnetic fields are known to bestow novel features
to electronic properties of the Weyl semimetals. Type-II WSM, in contrast to their type-I
counterpart, exhibits chiral anomaly only when the magnetic field is placed along the
direction of propagation of the particle. Chiral anomaly is absent when
magnetic field is perpendicular and the Landau level spectrum is gapped \cite{Soluyanov2015}.
Anomalous Hall conductivity depending on the tilt of the energy dispersions
and separation of the Weyl nodes, field-selective chiral
anomaly giving rise to a novel magneto-optical resonance has also been predicted \cite{Franz2013, *Zyuzin2016, *Udagawa2016}.
Coulomb interaction has been shown to restore the Lorentz symmetry in low-energy
regime in the 2D tilted Weyl semimetals subjected to both longitudinal
and transverse electromagnetic fields \cite{Isobe2016}.

Transport properties in presence of magnetic field has been investigated for
WSM type-I normal-superconductor junctions with spin-active interface.
In such junctions tunnelling conductance has been shown to depend on the direction of magnetic field
and the spin-flip scatterings are shown to influence the conductance spectra \cite{Asano2014}.
Magnetically induced edge states and their properties, transverse
conductance in a normal/superconducting WSM type I junction has been studied
for different superconducting pairing mechanisms \cite{Duan2018}.
Supression of Andreev reflection in a normal Weyl (type I) and $s$-wave superconductor
junction is shown to be lifted with a controlled Zeeman field \cite{Bovenzi2017}.
In Weyl type I junctions, perfect transmission rings are predicted due to resonance
of the Fermi vector signalling Klein tunnelling in presence of magnetic field \cite{Yesilyurt2016}.
For Weyl type-II $p-n$ junctions, novel quantum oscillations occurs due to momentum space Klein tunnelling in
presence of magnetic field \cite{OBrien2016}. In WSM/superconductor/WSM (type I)
heterostructures magnetic switching effect
is observed due to interplay of WSM surface spin polarizations and strong spin-orbit coupling
\cite{mohanta2019magnetic}. 
However, a detailed study of the influence of magnetic field on the transport properties
of junctions of WSM type-II with superconductors has not been undertaken so far.

In this work, we study the transport properties \textit{viz.}
tunnelling conductance of the Weyl-II semimetal junctions in presence of induced magnetization 
provided by proximity effect of thin magnetic film strips \cite{Mondal2010}.
We consider hetero-structure of normal, superconducting and magnetic Weyl semimetals.
The chemical potential in the different regions can be regulated by the use of an external gate voltage. 
Superconductivity is induced into WSM-II by extrinsic proximity effect
with an s-wave superconductor or with metallic point contacts \cite{Huang2018, *Li-Nano-2018, *Bachmanne2017} in the junction.
The induced magnetizations along $y,z$- directions does not destroy the topological
nature of the semimetal, only shifts the nodes in energy.
The topological properties of Weyl semimetals realised in
multilayer heterostructures of normal insulators and topological insulators
are shown to survive in the presence of both inversion and time reversal
symmetry breaking terms, provided the inversion symmetry breaking is not strong \cite{Burkov2012}.
However, we note here that the effect is distinct from that considering finite magnetization
along the $x$ direction which gaps out the spectrum and Weyl nodes are destroyed.
We study the effect of induced magnetization 
on the differential conductance of a normal/magnetic/superconducting (NMS) junction 
based on time-reversal symmetric type II WSM. We find that in the normal (magnetic) junction with superconductor, the Andreev reflected
modes exhibit both spin singlet and spin triplet pairing. The back scattering of electrons from the magnetic-superconducting
interface is prohibited due to the unavailability of channels supporting the process.
The differential tunnelling conductance is found to be independent of the strength of the magnetization 
and the width of the magnetic barrier.

The organization of the rest of the paper is as follows. In Sec. \ref{NMS}, we detail out the model for a
heterostructure of normal/magnetic/superconducting Weyl semimetal and calculate the tunnelling
properties and subgap conductance for the junction. We proceed with providing analytical calculations of the 
otherwise complicated Andreev reflection coefficients in the thin barrier limit and for the normal incidence.
We discuss possible experiments to test our theoretical predictions in Sec.\ref{Disc}.

\section{N-M-S junction}
\label{NMS}

In this section we study the conductance through a N-M-S junction, where $N$ denotes
a normal Weyl-II semimetal region, $M$- a Weyl-II semimetal where the proximity induced field 
has been induced across a width $L$, and $S$ denoting
Weyl-II region where superconductivity is induced by proximity effect with a conventional $s$-wave superconductor.
The junction is described by the Bogoliubov de-Gennes (BdG) Hamiltonian which is of the form:
\begin{equation}\label{BdGH}
\mathcal{H}_{BdG}
=\begin{pmatrix}
H_0(x) + \mathcal{M}(y,z)   & \Delta(x) \\
\Delta^{*}(x) & -H_0(x) + \mathcal{M}(y,z)
\end{pmatrix}
\end{equation}
where $H_0(x) = \hbar v_1 p_x \sigma_0 + \sgn(v_2) \hbar v_2 \bm {\mathrm{p . \sigma}}-\mu(x)\Theta(x)-\mu_s\Theta(L+x)$,
$\mathcal{M}(y,z)=M_{y}\sigma_{y}[\Theta(x)-\Theta(x-L)]-M_{z}\sigma_{z}[\Theta(x)-\Theta(x-L)]$ is the exchange energy and $M_{y(z)}$
is it's magnitude along $y(z)$ axis,
$\Theta(x)$ denotes the Heaviside step function, $v_1$ is the spectrum tilt, $v_2$ is the Fermi velocity
and $\sgn (v_2)=\pm$ denotes the chirality of the nodes, ${\bf \sigma}$ are the Pauli matrices, $\mu(x)$ is the chemical
potential such that $\mu(x)= \mu, \; x<0$ and the pairing potential, $\Delta(x)=\Delta_0 e^{i \phi}$ for region $x>L$,
$\mu_s$ is electrostatic potential controlled by an external gate voltage, $\mu_s\gg\Delta_0,\mu(x)$.
For type-II Weyl semimetals, $v_1>v_2$.  In the magnetic region, the
magnetic strips are placed both at the top and bottom of the region.
$\mathcal{M}(y,z)$ is analogous to the vector potential arising due to a
 magnetic field ${\bf B_{\alpha}}= M_{\alpha^{\prime}}/e v_2 [\delta(x) -\delta(x-L)]\hat{\alpha}^{\prime}$,
$\hat{\alpha}^{\prime} = \hat{z} (\alpha= y), \hat{y} (\alpha=z)$. 
The superconducting pairing potential couples electrons and holes
near time-reversed Weyl nodes with same chirality. 
The BdG Hamiltonian in Eqn. \ref{BdGH} can be diagonalised in the
$\Phi=(c_{1,\uparrow}, c_{1,\downarrow},c^{\dagger}_{2,\downarrow},c^{\dagger}_{2,\uparrow})^{T}$
basis for different regions as described below.
\begin{figure}[!ht]
\includegraphics[width=.5\linewidth]{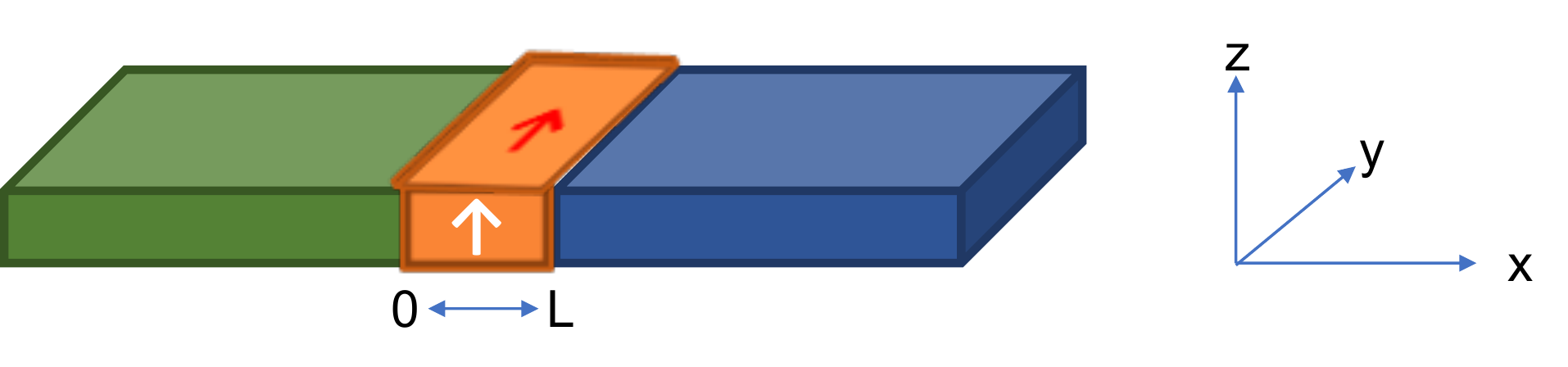}
\caption{Schematic representation of N-M-S junction with magnetic strips on top and side surfaces.
The green region denotes the normal WSM-II, orange denotes the magnetic region, while the blue denotes
the superconducting part of the junction. The magnetic strips induces the desired magnetization
in the region $0 \leq x \leq L$. The red and white arrows denotes the direction of induced magnetization along  $\hat{y}$ and $\hat{z}$ respectively. }
\label{fig1}
\end{figure}
In the normal region, the electron and hole energy dispersions are:
\begin{eqnarray}
 E_{e\pm}&=\hbar v_1 p_x \pm \hbar v_2 |\bm{p}|-\mu, \quad 
 E_{h\pm}&=-\hbar v_1 p_x \pm \hbar v_2 |\bm{p}|+\mu,
\end{eqnarray} 
with $|\bm{p}|=\sqrt{p^2_x+p^2_{\pr}},\;\;p_{\pr}=\sqrt{p^2_y+p^2_z}$.
At a given incident energy $E$ with conserved transverse
momenta $p_{\pr}$, the eigenstates are :
\begin{align}\label{wavefnn}
\Psi_{e\pm} &=
\begin{pmatrix}
\pm p_{\pm} + p_z, & p_{x \pm} + i p_y, & 0, & 0 \end{pmatrix}^T
e^{i( p_{x\pm}x+ p_y y + p_z z)}, \nonumber \\
\Psi_{h \pm} &=
\begin{pmatrix}
0, &0, &\mp p^{'}_{\pm} +p_z, &p^{'}_{x \pm} + i p_y \end{pmatrix}^T
e^{i( p^{'}_{x \pm}x+ p_y y + p_z z)},
\end{align}
\noindent
with,
$p_{x \pm} =\frac{v_1 (E+\mu) \mp v_2 \sqrt{(E+\mu)^2 +\hbar^2 p^2_{\pr}(v^2_1-v^2_2)}}{\hbar(v^2_1- v^2_2)}$ for electrons,
$p^{'}_{x \pm} = \frac{ v_1 (\mu-E) \pm v_2 \sqrt{(E-\mu)^2 +\hbar^2 p^2_{\pr}(v^2_1-v^2_2)}}{\hbar(v^2_1- v^2_2)}$ for holes,
$p_{\pm}=\sqrt{({p_{x \pm}})^2+p^2_{\pr}}$, $p^{'}_{\pm} =\sqrt{({p^{'}_{x \pm}})^2+p^2_{\pr}}$.
In the magnetic region,
\begin{align}
 E^M_{e\pm}=\hbar v_1 p^m_x \pm \hbar v_2 |\bm{p^m}|, \quad 
E^M_{h\pm}=-\hbar v_1 p^{'m}_x \pm \hbar v_2 |\bm{p^{'m}}|,
\end{align}
are the dispersion of the electron and holes respectively.
Here, $|\bm{p^m}|={p^m}=\sqrt{(p^m_{x\pm})^2+\tilde{p}^{2}_{e,y}+\tilde{p}^{2}_{e,z}},\;\;|\bm{p^{'m}}|=p^{'m}=\sqrt{(p^{'m}_{x\pm})^2
+\tilde{p}^{2}_{h,y}+\tilde{p}^{2}_{h,z}}$, $\tilde{p}_{(e/h),y}=p_y\pm \mathcal{B}_{y}$, $\tilde{p}_{(e/h),z}=p_z\mp \mathcal{B}_{z}$,
$\mathcal{B}_{\alpha}=M_{\alpha}/\hbar v_2$ is the induced magnetization.
The eigenstates in the magnetic region at a given incident energy $E$ are:
\begin{align} \label{wavefnm1}
\Psi^{m}_{e\pm} &=
\begin{pmatrix}
\pm p^m_{\pm} +\tilde{p}_{e,z}, & p^m_{x \pm} + i \tilde{p}_{e,y}, & 0, & 0 \end{pmatrix}^T
e^{i( p^m_{x\pm}x+ p_y y + p_z z)}, \nonumber \\
\Psi^{m}_{h \pm} &=
\begin{pmatrix}
0, &0, &\mp p^{'m}_{\pm} +\tilde{p}_{h,z}, &p^{'m}_{x \pm} + i \tilde{p}_{h,y} \end{pmatrix}^T
e^{i( p^{'m}_{x \pm}x+ p_y y + p_z z)}
\end{align}
with the longitudinal momenta for the electrons $p^m_{x \pm}$
$p^m_{x \pm} =
\frac{ v_1 E \mp v_2 \sqrt{E^2+ \hbar^2 (\tilde{p}^2_{e,y}+\tilde{p}^2_{e,z})(v^2_1-v^2_2)}}{\hbar(v^2_1- v^2_2)}$, and
holes $p^{'m}_{x \pm} =
\frac{ - v_1 E \pm v_2 \sqrt{E^2+ \hbar^2 (\tilde{p}^2_{h,y}+\tilde{p}^2_{h,z})(v^2_1-v^2_2)}}{\hbar(v^2_1- v^2_2)}$.
The excitation energy for the electron and hole like quasiparticles in the superconductor is:
\begin{equation}\label{ESup}
 E_{\pm}=\sqrt{\Delta^2_0 + (\hbar v_1 p^s_x \pm \hbar v_2 |\bm{p}| -\mu_s)^2},
\end{equation}
with $|\bm{p}|=\sqrt{p^{s 2}_x+p^2_{\pr}}$, $p^s_x=k_{1(2)} \simeq \frac{\mu_s}{\hbar(v_1\pm v_2)}$ for $\Psi^S_{\pm}$.
Considering $\mu_s\gg(\Delta_0,\mu(x),E)$, the outgoing eigenstates in the superconducting region are:
\begin{align}\label{wavefnso}
\Psi^{S,o}_{+} &= \begin{pmatrix} e^{i \beta},& e^{i \beta}, &1, &1 \end{pmatrix}^T e^{ i(k_1 x + p_y y + p_z z)-\tau_1 x},\nonumber \\
\Psi^{S,o}_{-} &= \begin{pmatrix} e^{i \beta}, & -e^{i \beta}, &1, &-1 \end{pmatrix}^T e^{ i(k_2 x + p_y y + p_z z)-\tau_2 x}
\end{align}
while the incoming states are:
\begin{align}\label{wavefnsi}
\Psi^{S,i}_{+} &= \begin{pmatrix} e^{-i \beta}, &e^{-i \beta}, &1, &1 \end{pmatrix}^T e^{ i(k_1 x + p_y y + p_z z)+\tau_1 x},\nonumber \\
\Psi^{S,i}_{-} &= \begin{pmatrix} e^{-i \beta},&-e^{-i \beta}, &1, &-1 \end{pmatrix}^T e^{ i(k_2 x + p_y y + p_z z)+\tau_2 x},
\end{align}

where, $\tau_1 \simeq \frac{\Delta_0 \sin \beta}{\hbar(v_1+v_2)}, \quad \tau_2 \simeq \frac{\Delta_0 \sin \beta}{\hbar(v_1-v_2)}$ are
the localization lengths for $\Psi^{S,i(o)}_+$ and $\Psi^{S,i(o)}_-$ respectively, and
$\beta=\arccos(E / \Delta_0)\; \Theta(\Delta_0-E)
-i \arccosh(E / \Delta_0)\; \Theta(E-\Delta_0)$.

\subsection{Conductance}
\label{NMS-C}

We consider $\Psi_{e -(+)}$ as the initial incident mode and observe that the initial incident mode can also be
a linear superposition of the $e_{\pm}$ modes.
To compute the tunnelling conductance we assume clean interfaces, match the boundary conditions,
and subsequently retrieve the Andreev reflection coefficients. Note that the
condition of matching of the derivatives of the wavefunctions at the boundaries
is not applicable for this junction. Hence, the limit of thin barrier with barrier width $L\rightarrow 0$
and  strength of induced magnetization $\mathcal{B} \rightarrow \infty$
is not applicable at the onset \cite {BTK82}.
In the normal WSM (I) region ($x<0$), the total wave function is:
\begin{align}\label{dr9}
\varPhi_{I}(\bm{r}) & = \Psi_{e \varmp} + r_1 \Psi_{h \varmp} + r_2 \Psi_{h \varpm},
\end{align}
where $r^{\mp}_1$ and $r^{\pm}_2$ are the retro and specular Andreev reflection amplitudes for $e-(+)$
incident modes respectively.
\begin{align}
\varPhi_{II}(\bm{r})=c_1\Psi^{m}_{e +} +c_2\Psi^{m}_{e -} + c_3\Psi^{m}_{h +} + c_4\Psi^{m}_{h -},
\end{align}
is the wave function in the magnetic region, with $c_{1(2)}$ and $c_{3(4)}$ are the amplitudes
of the electron($+(-)$) and hole ($+(-)$) modes in the magnetic barrier region.
While in the superconducting region the total wavefunction is of the form,
\begin{align}
\Phi_{III}(\bm{r}) &= a_1 \Psi^{S,o}_{+} + a_2 \Psi^{S,o}_{-},
\end{align}
where $a_1$ and $a_2$ are the amplitudes of the right moving electron and hole like quasiparticles.
Matching the wavefunctions at the two different
interfaces in the junctions we get, 
\begin{align} \label {bc}
\varPhi_{I}(\bm{r})|_{x=0^{-}}&=\varPhi_{II}(\bm{r})|_{x=0^{+}}, \nonumber \\
\varPhi_{II}(\bm{r})|_{x=L^{-}}&=\varPhi_{III}(\bm{r})|_{x=L^+}
\end{align}
For the finite barrier we note that compact analytical forms of the reflection
coefficients indicating the dependence of magnetic field is not easily feasible. However, 
in the following sections, we show that in the limiting cases for normal incidence on a finite
barrier and for the thin barrier limit such analytical forms can be obtained. 
The retro and specular Andreev reflection coefficients are scaled in terms of the reflected hole
current density normalized to the incident electron current density \cite{BTK82}
and hence, can be obtained as:
\begin{align}\label{dr13}
 A^{\pm}_{1(2)}= \left|\frac{\langle\Psi_{h\pm}|J_x|\Psi_{h\pm}\rangle}{\langle\Psi_{e\pm}|J_x|\Psi_{e\pm}\rangle}\right||r_{1(2)}|^2
\end{align}
where $J_x = 1/\hbar [x,\mathcal{H}_{BdG}]$ is the current density operator along $\hat{x}$ direction.
\begin{figure}
\includegraphics[width=0.4\textwidth]{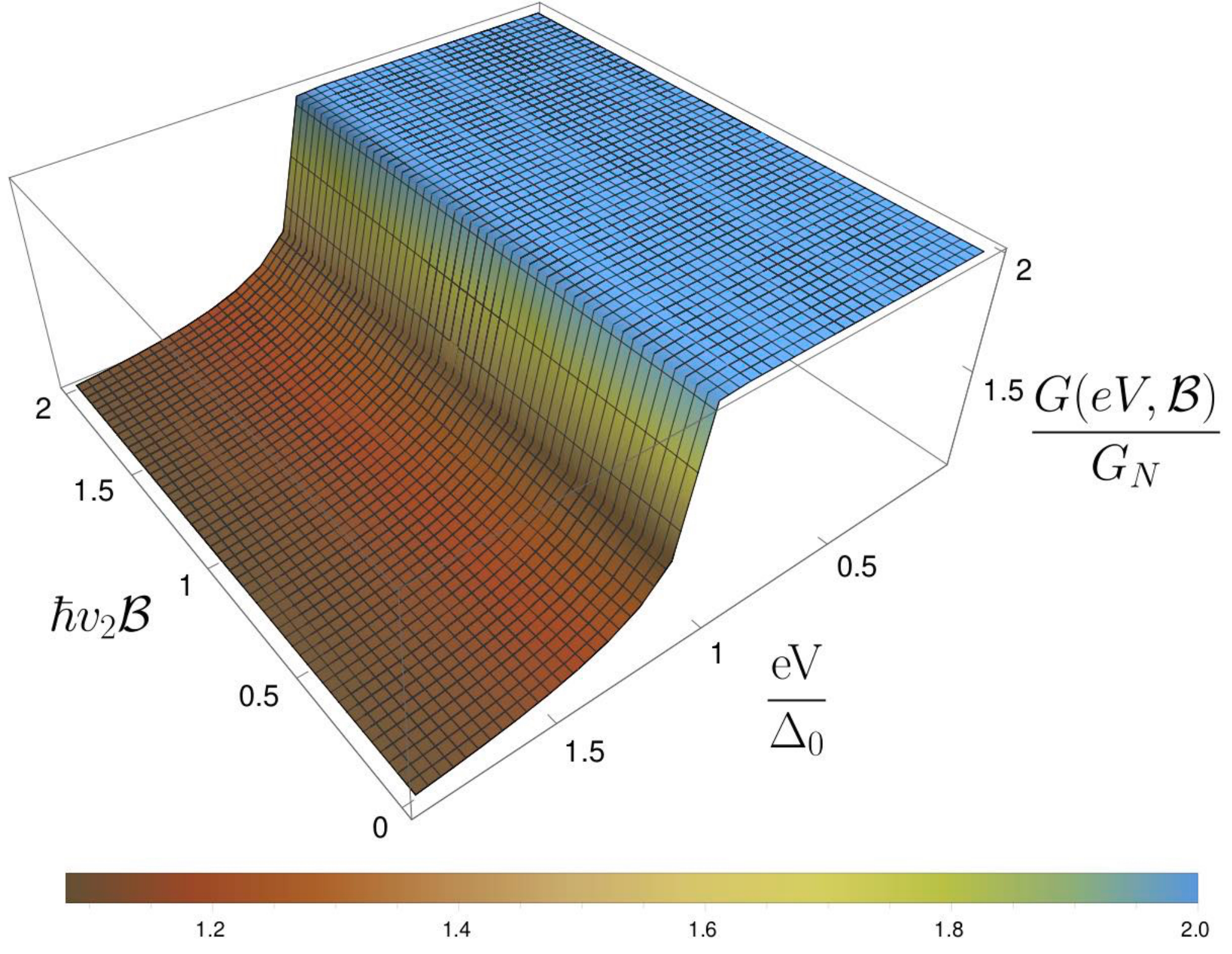}
\caption{Differential conductance as function of bias voltage and induced magnetization.
$v_1/v_2$ in normal/magnetic/superconducting
region =2, $\mu=0.5 \Delta_0$ for normal and magnetic region, $\mu_s=100 \Delta_0$ in the superconducting region.}
\label{fig2}
\end{figure}
\begin{figure}[ht]
\includegraphics[width=0.35\textwidth]{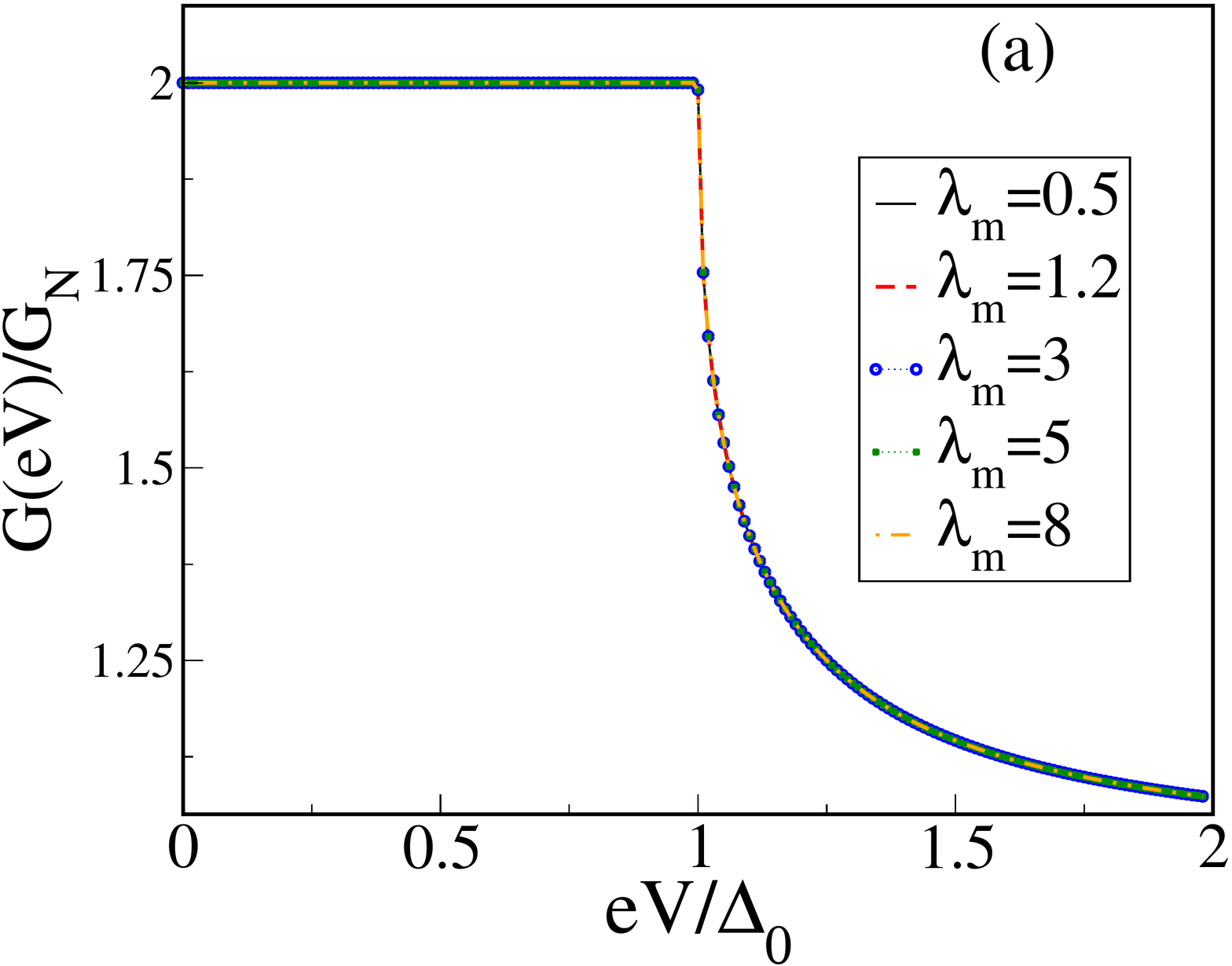} \hspace{1.5cm}
\includegraphics[width=0.35\textwidth]{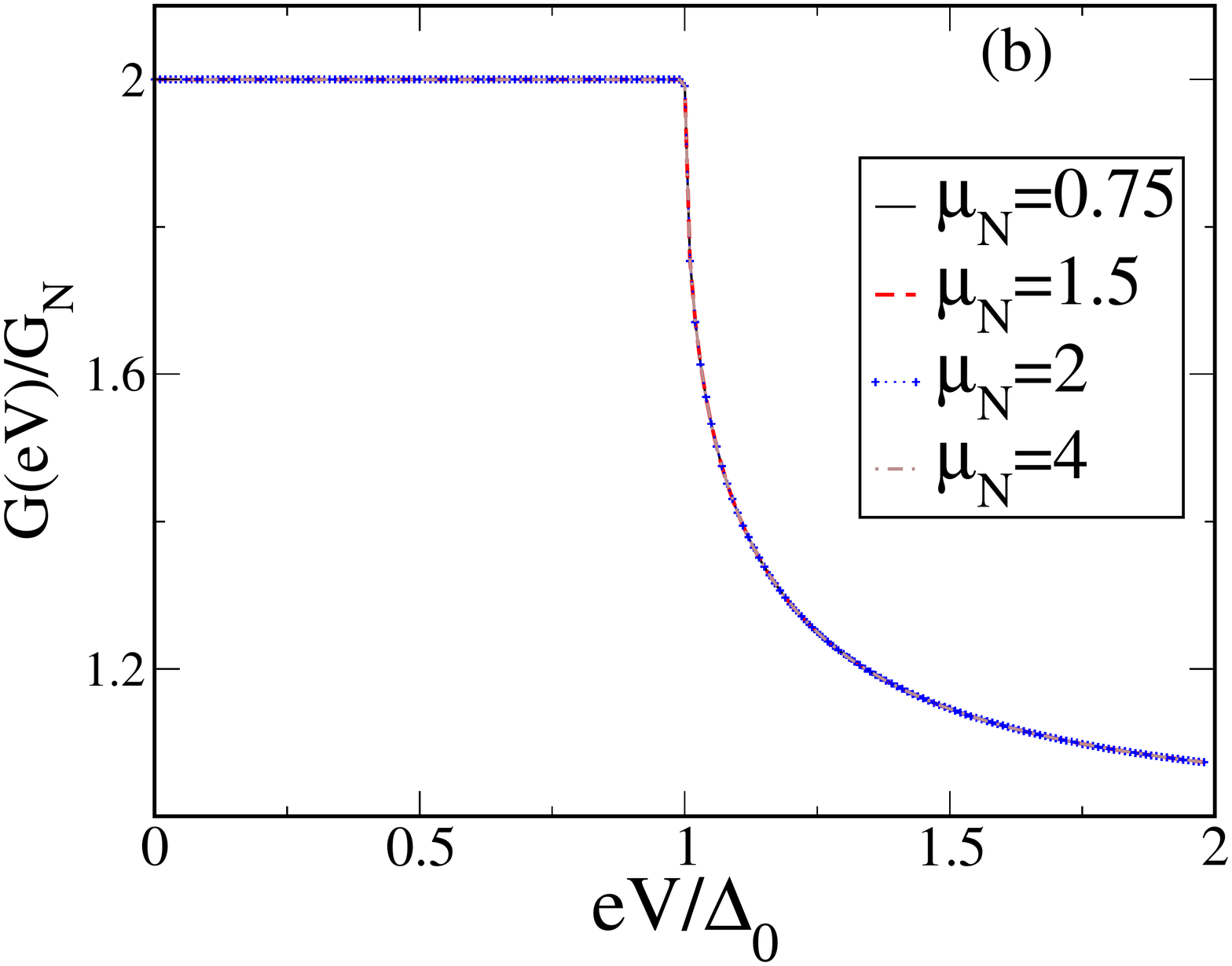}
\caption{(a)Differential conductance as function of bias voltage for different representative value of
magnetic barrier width $\lambda_m$, $v_1/v_2$ in normal/magnetic/superconducting
region =$2/\sqrt{3}$, $\mu=0.5$ for normal and magnetic region, $\mathcal{B}_{y/z}=2$.
 (b) Differential conductance as function of bias voltage for different representative value of
chemical potential in the normal region
$\mu_N$ scaled in terms of $\Delta_0$, $\lambda_m=4$, $\mathcal{B}_{y/z}=2$.}
\label{fig3}
\end{figure}
The differential conductance of the junction can be derived as:
\begin{align}\label{defG}
G(eV) = \frac{e^2 S}{4\pi^2 h}\sum_j\int d^2 \mathbf{p}_{\pr} [1+ A^{\pm}_j(\mathbf{p}_{\pr},eV) - R^j_N(\mathbf{p}_{\pr},eV)], 
\end{align}
where $S$ is the cross sectional area of the junction, $j=1,2$
corresponding to the $e+$ and $e-$ incident modes for Weyl nodes at $K_0(-K_0)$
(of same chirality) and $R^{j}_N$ are the normal electron reflection coefficients.
Due to the titled spectrum the electron back scattering channels are absent. Hence the normal reflection coefficient
for this junction $R^{j}_N=0$.
$G(eV)$ receives equal contribution from Weyl nodes of opposite chirality. The differential conductance normalized
to $G_N(eV)= \frac{e^2 S }{2\pi h}\mathcal{N}(eV)$, corresponding to the number of
available channels with energy $(eV+\mu_N)$ on the normal side is:
\begin{align}\label{Condf}
G(eV)/G_N(eV) = \frac{1}{2\pi \mathcal{N}(eV)} 
\sum_{j,\alpha=\pm K_0}\int d^2 \mathbf{p}_{\pr} [1+ A^{\pm,\alpha}_j(\mathbf{p}_{\pr},eV)].
\end{align}
where $\mathcal{N}(eV)=((eV+\mu_N)\sin \alpha_c)^2$, $\alpha_c=\arctan v_2/\sqrt{v_1^2-v_2^2}$
is the critical angle of incidence. The cut-off for transverse momenta is bounded by allowed channels
$p^{max}_{\pr} =\sqrt{\mathcal{N}(eV)}$ on the interface.

\subsection{Results}
\subsubsection{Finite barrier}
\label{Fb1}
We plot the differential conductance as expressed in Eqn.(\ref{Condf})
as a function of bias voltage and induced magnetization in Fig.\ref{fig2}.
The differential conductance is found to be robust and independent of the strength of the magnetization
within the sub-gap and decreases monotonically as the bias increases. To understand this robust behaviour,
consider particles incident on the interface of the magnetic-superconducting region
in the $x-y$ plane. For this case, the sum of the Andreev reflection coefficients can be shown to be dependent only on
the gap energy and the energy of the incident particles, and independent of any other energy
scales in the system. In particular for a given incident energy $E$,  $ A_1+A_2= \Theta(E-\Delta)+
\frac{1}{\Delta^2}(E-\sqrt{E^2-\Delta^2})^2 \Theta(\Delta-E)$. We provide an analytical proof for 
the normal incidence in Sec.\ref{Fb2}.
However, the differential conductance receives contribution
from particles incident on the interface from all possible allowed channels.
Hence, it is difficult to single out the contribution from the normal incidence in the total conductance.
To this end, we look into the spin configuration of the incoming electrons from the conduction ($e+$)
and valence ($e-$) bands in the normal and magnetic region.
The spin expectation values $<S>=\hbar/2 \hat{\sigma}$ are calculated for each of the
above branches. 
The spin expectation value in the normal and magnetic region in units of $\hbar/2$ is given by:

\begin{align}
 <S_{e\pm}> & = \Big[2 p_{x \pm}(\pm p_{\pm}+p_z)/\mathcal{N}^{\pm}_e, 2 p_{y}(\pm p_{\pm}+p_z)/\mathcal{N}^{\pm}_e,  
 2 p_{z}(\pm p_{\pm}+p_z)/\mathcal{N}^{\pm}_e \Big], \nonumber \\
  <S^M_{e\pm}> & = \Big[2 p^m_{x \pm}(\pm p^m_{\pm}+\tilde{p}_{e,z})/\mathcal{N}^{\pm,M}_e, 
  2 \tilde{p}_{e,y}(\pm p^m_{\pm}+\tilde{p}_{e,z})/\mathcal{N}^{\pm,M}_e, 
 2 \tilde{p}_{e,z}(\pm p^m_{\pm}+\tilde{p}_{e,z})/\mathcal{N}^{\pm,M}_e \Big], \nonumber \\
 <S_{h\pm}> & = \Big[2 p^{\prime}_{x \pm}(\mp p^{\prime}_{\pm}+p_z)/\mathcal{N}^{\pm}_h, 2 p_{y}(\mp p^{\prime}_{\pm}+p_z)/\mathcal{N}^{\pm}_h, 
 2 p_{z}(\mp p^{\prime}_{\pm}+p_z)/\mathcal{N}^{\pm}_h \Big], \nonumber \\
  <S^M_{h\pm}> & = \Big[2 p^{\prime m}_{x \pm}(\mp p^{\prime m}_{\pm}+\tilde{p}_{e,z})/\mathcal{N}^{\pm,M}_h,
  2 \tilde{p}_{e,y}(\mp p^{\prime m}_{\pm}+\tilde{p}_{e,z})/\mathcal{N}^{\pm,M}_h,
  2 \tilde{p}_{e,z}(\mp p^{\prime m}_{\pm}+\tilde{p}_{e,z})/\mathcal{N}^{\pm,M}_h \Big],
\end{align}

where the different momenta of the incoming particles in different region has been defined previously and
$\mathcal{N}^{\pm}_e= \sqrt{2 p_{\pm}(\pm p_{\pm}+p_z)}$, $\mathcal{N}^{\pm,M}_e= \sqrt{2 p^m_{\pm}(\pm p^m_{\pm}+\tilde{p}_{e,z})}$
$\mathcal{N}^{\pm}_h= \sqrt{2 p^{\prime}_{\pm}(\mp p^{\prime}_{\pm}+p_z)}$, and $\mathcal{N}^{\pm,M}_h= \sqrt{2 p^{\prime m}_{\pm}(\mp p^{\prime m}_{\pm}+\tilde{p}_{e,z})}$
are the normalization of the wavevectors in the normal and magnetic region respectively. 

\begin{figure*}
 \includegraphics[width=0.227\textwidth]{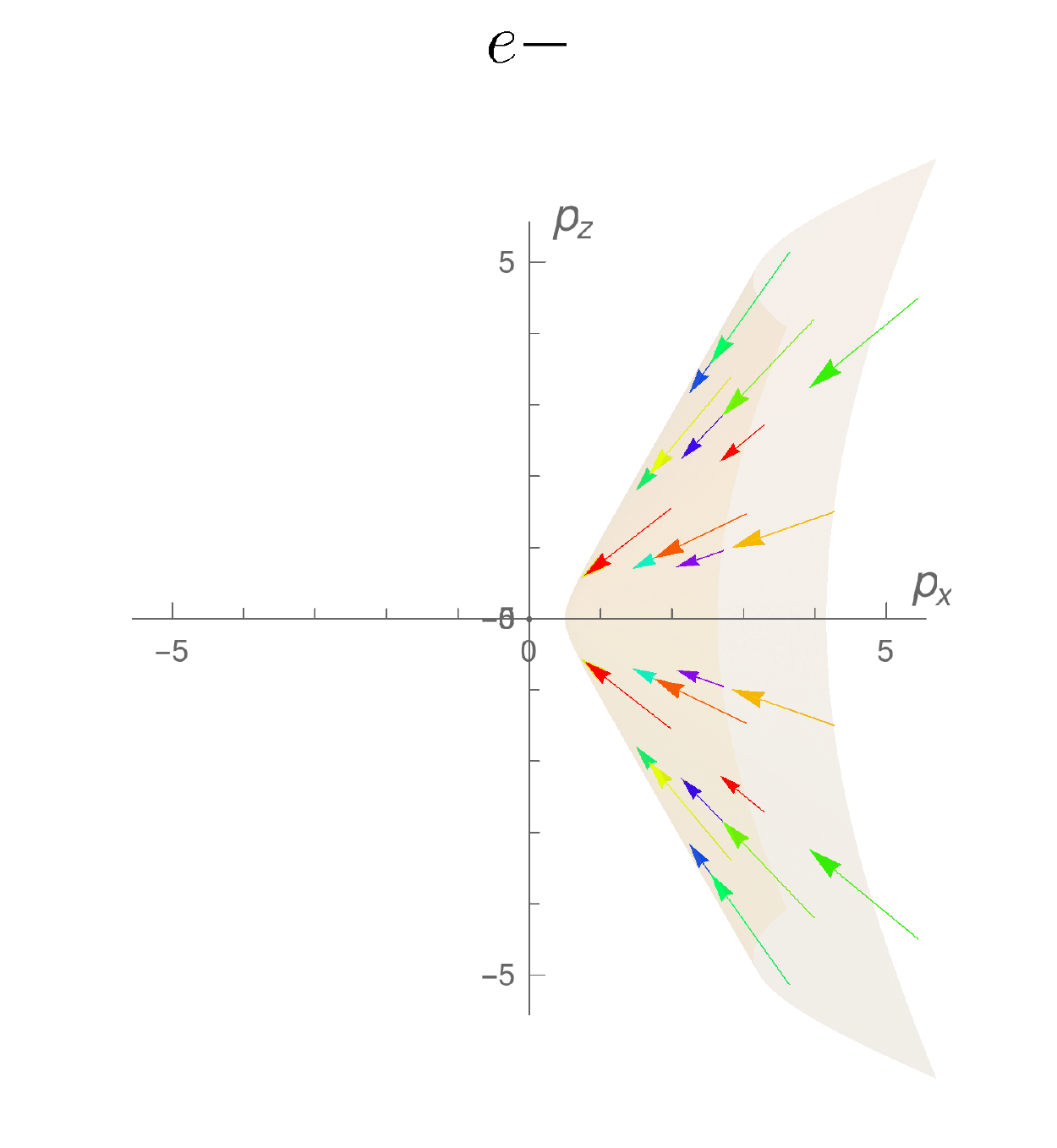}
 \hspace{.2cm}
 \includegraphics[width=0.227\textwidth]{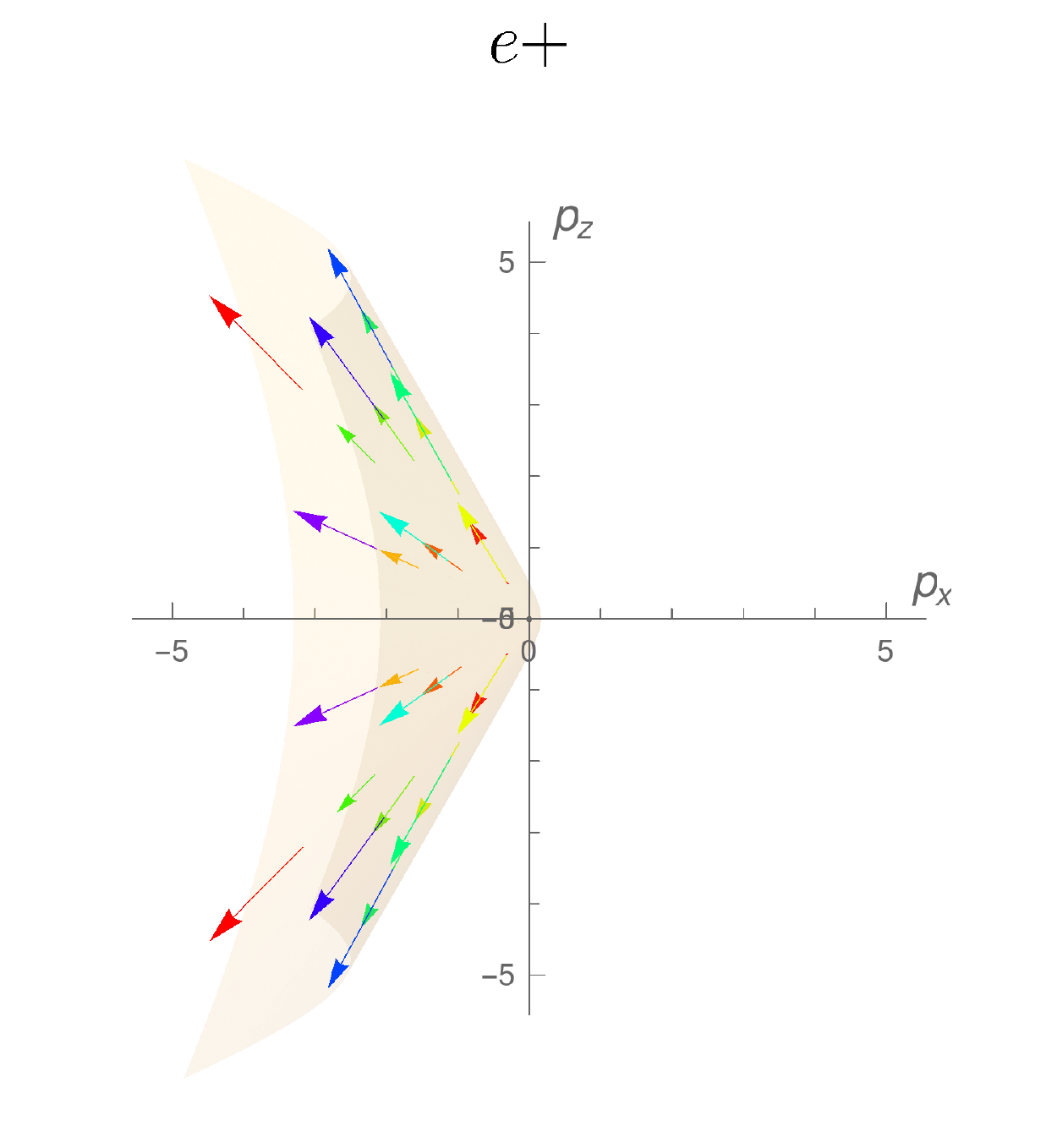}
  \hspace{.2cm}
  \includegraphics[width=0.227\textwidth]{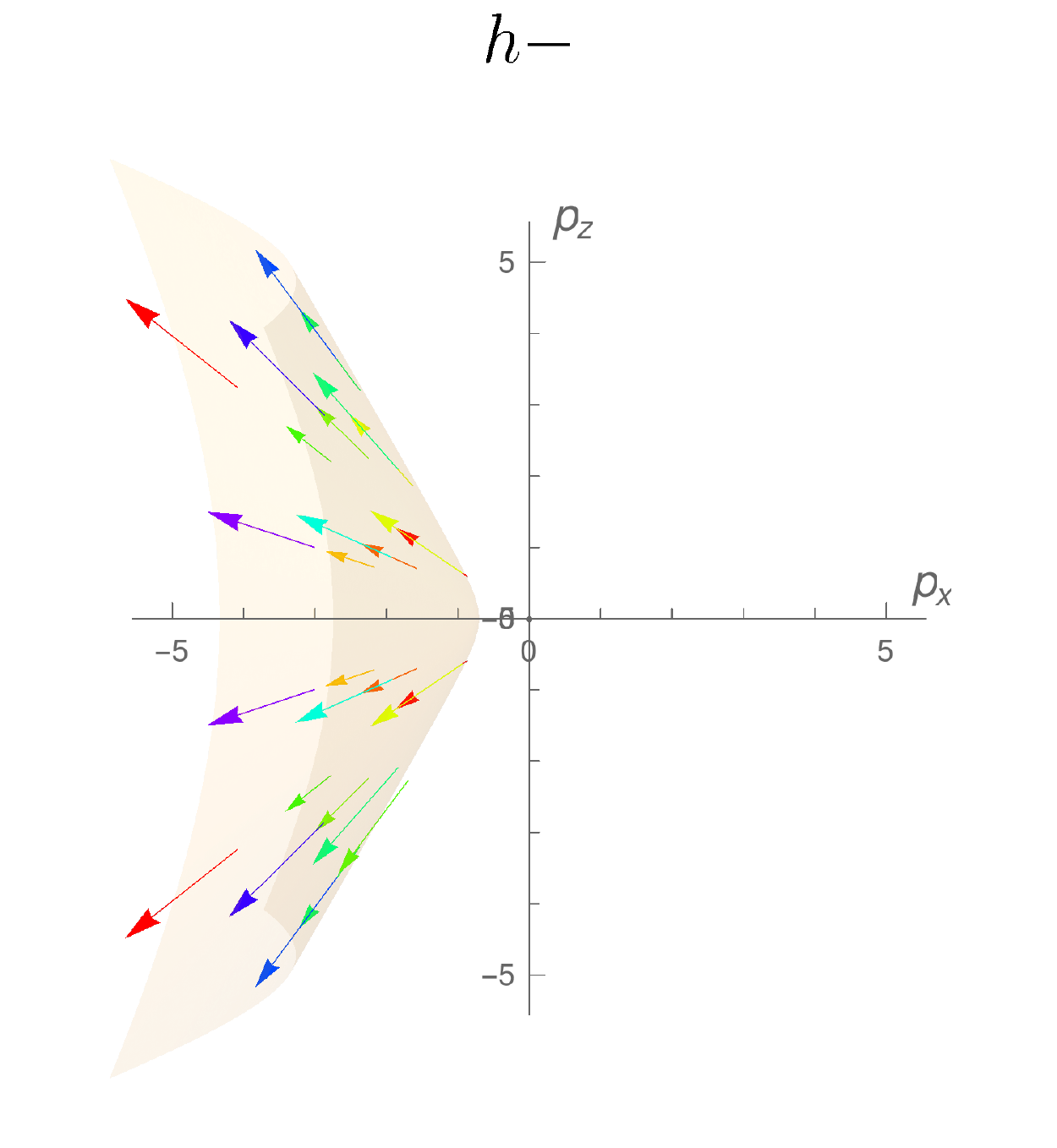}
 \hspace{.2cm}
 \includegraphics[width=0.227\textwidth]{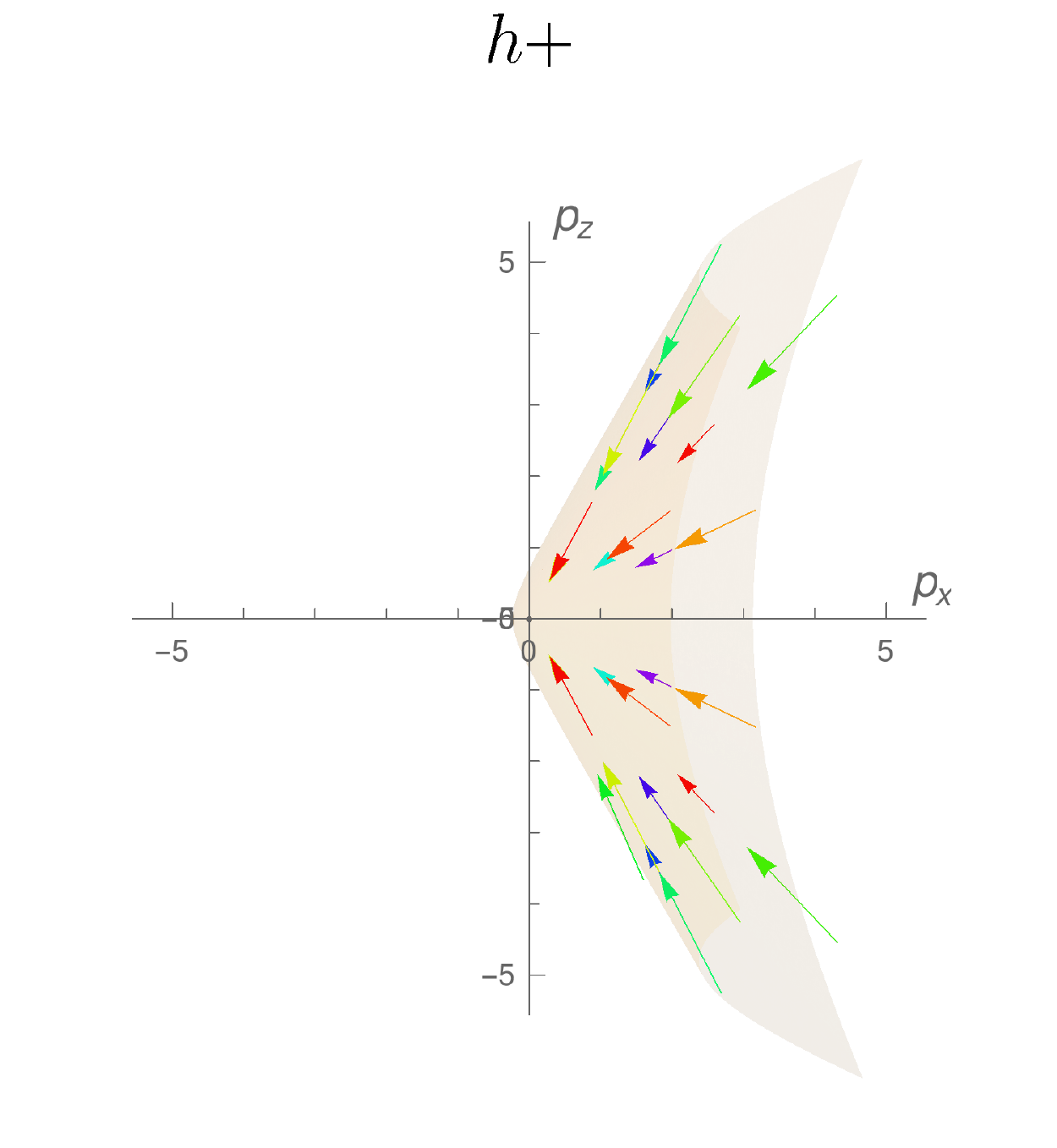}\\
 \includegraphics[width=0.227\textwidth]{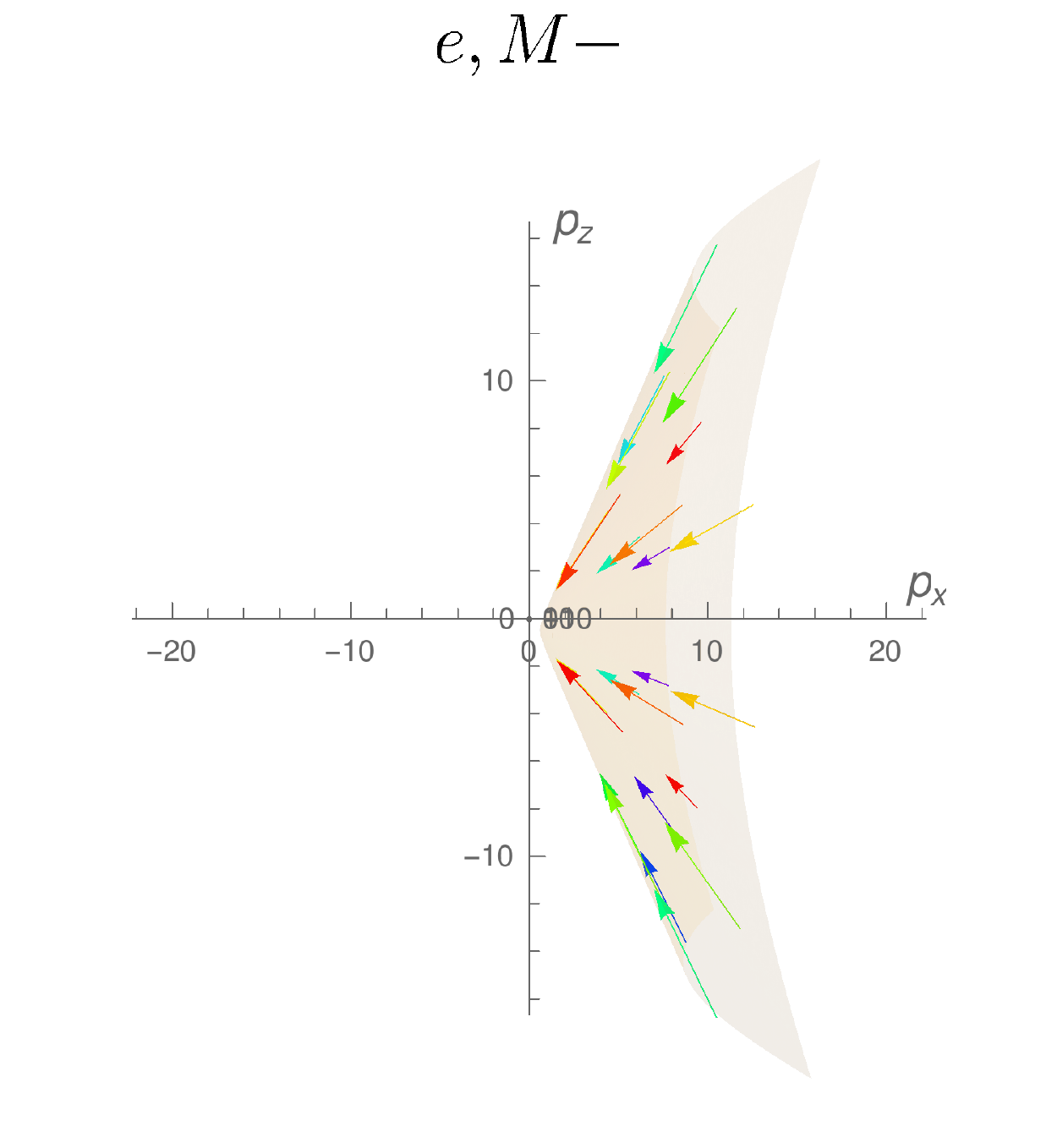}
 \hspace{.3cm}
 \includegraphics[width=0.227\textwidth]{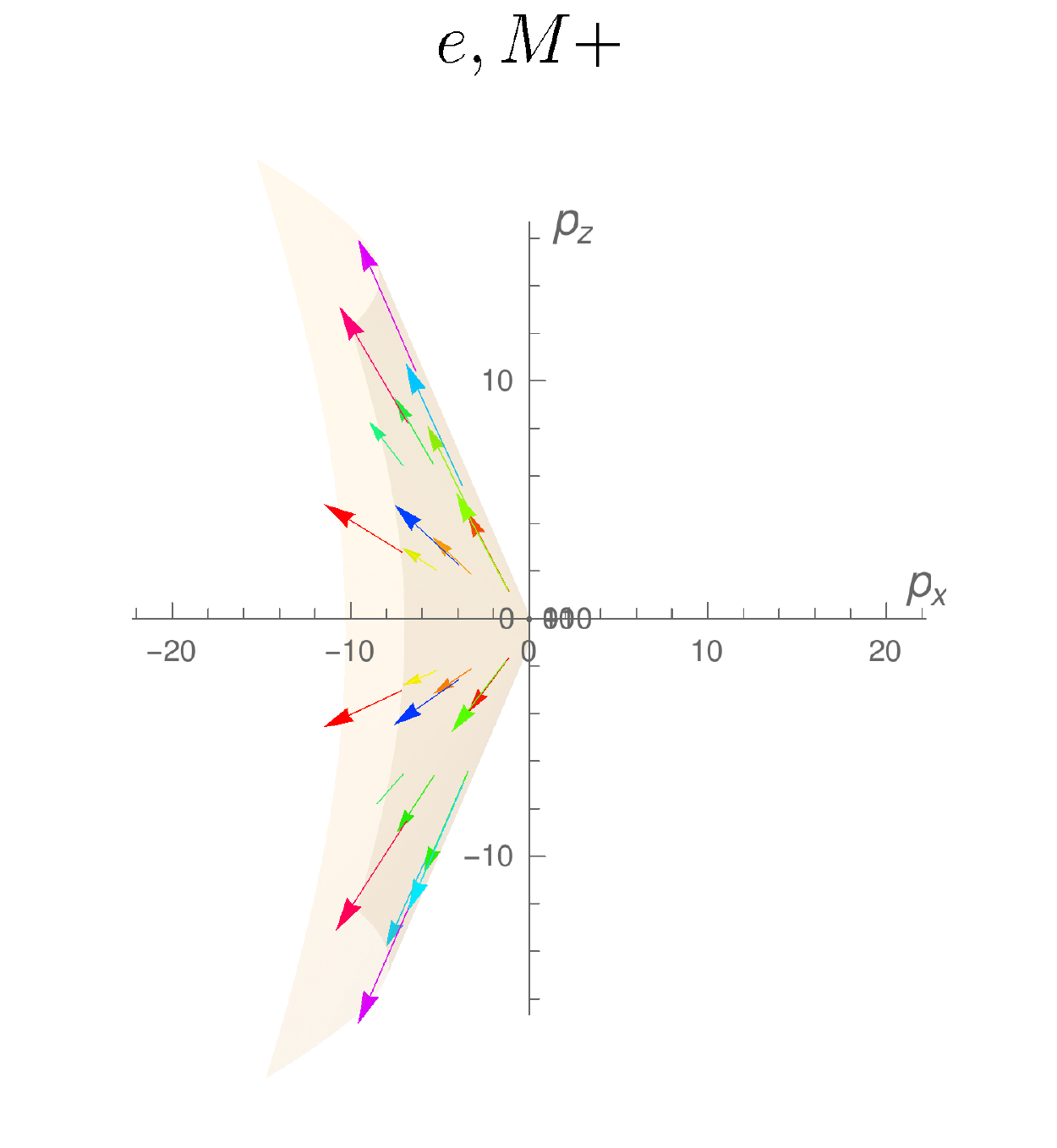}
  \hspace{.3cm}
  \includegraphics[width=0.227\textwidth]{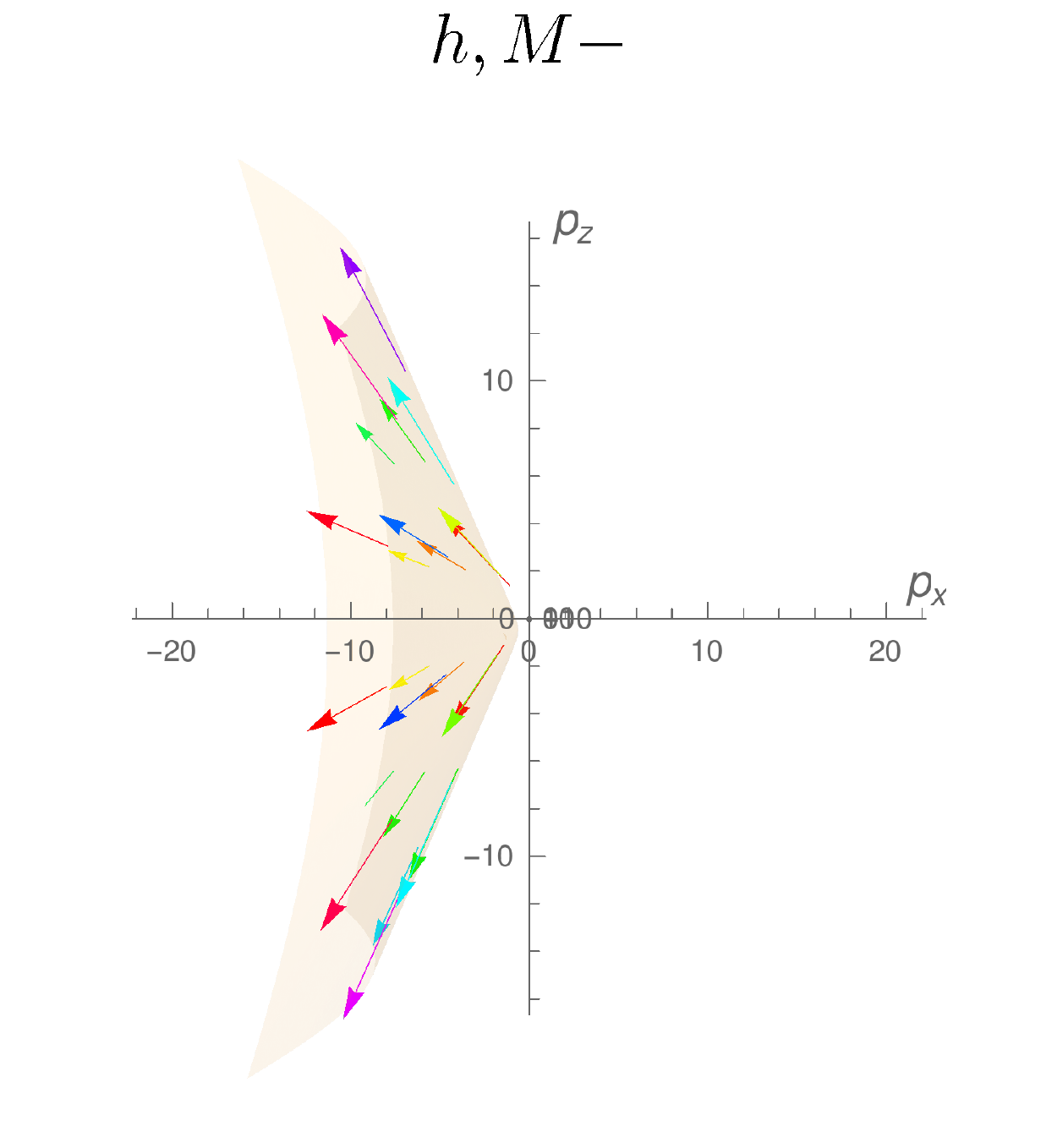}
 \hspace{.3cm}
 \includegraphics[width=0.227\textwidth]{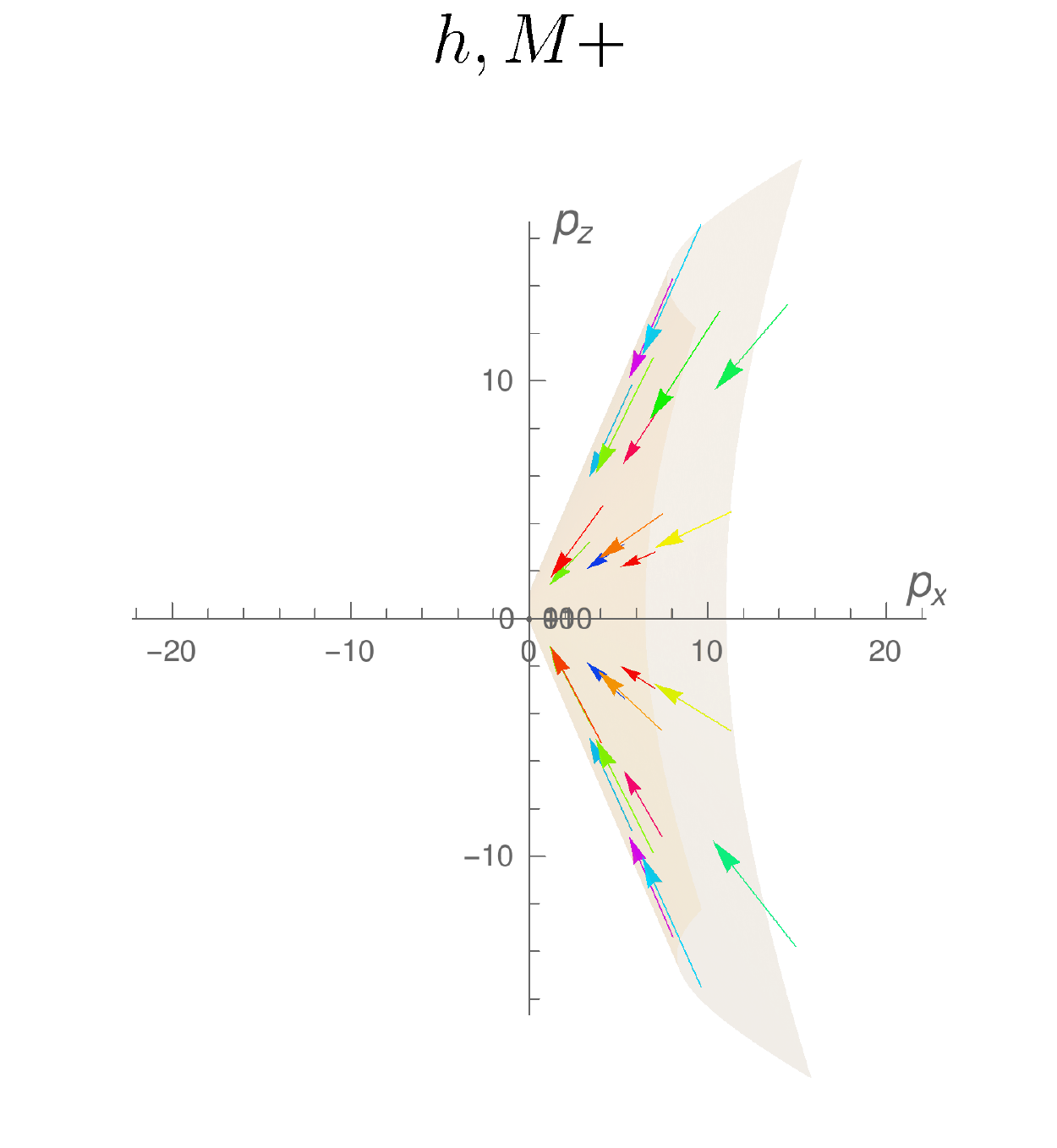}
  \includegraphics[width=0.227\textwidth]{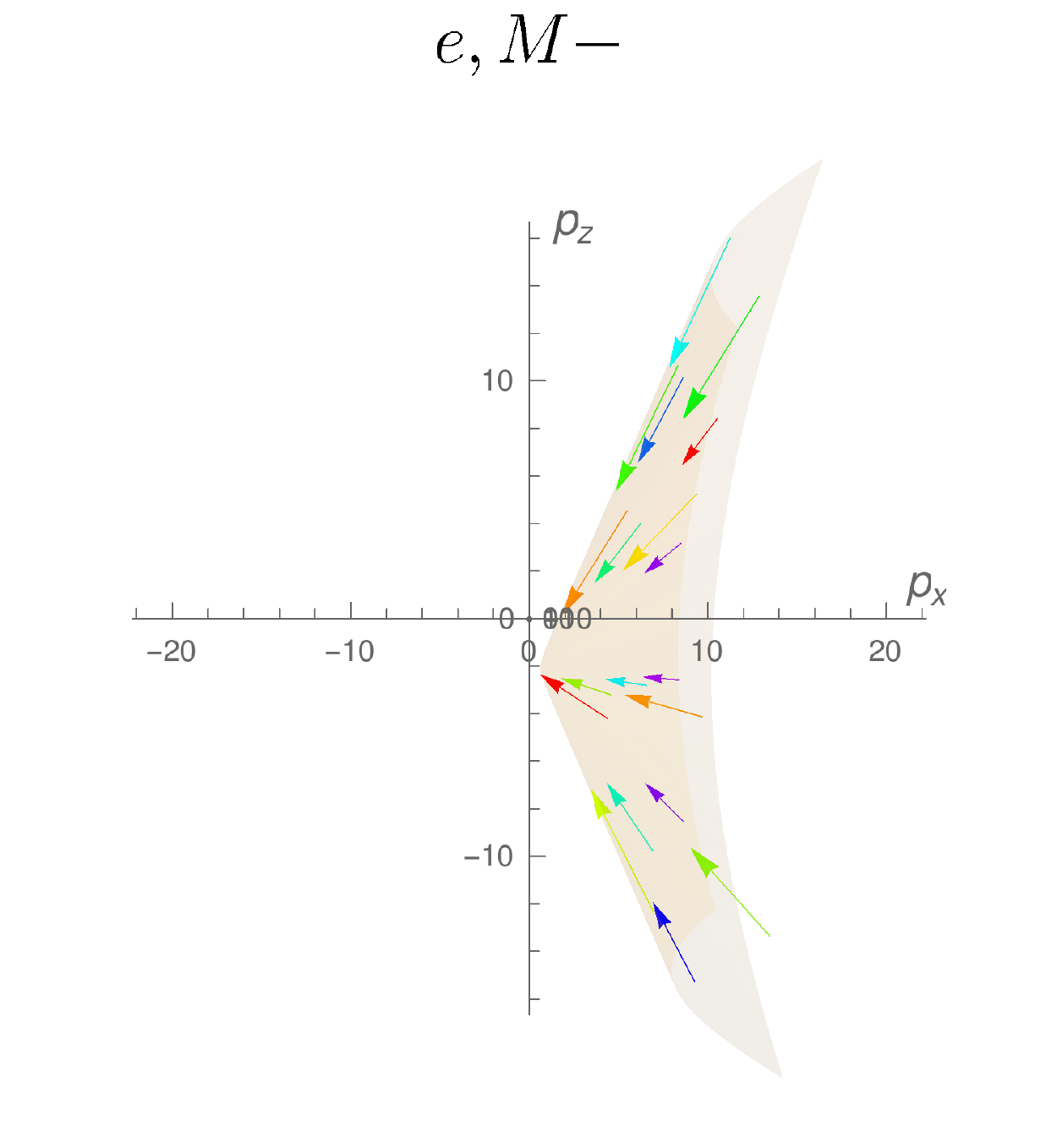}
 \hspace{.3cm}
 \includegraphics[width=0.227\textwidth]{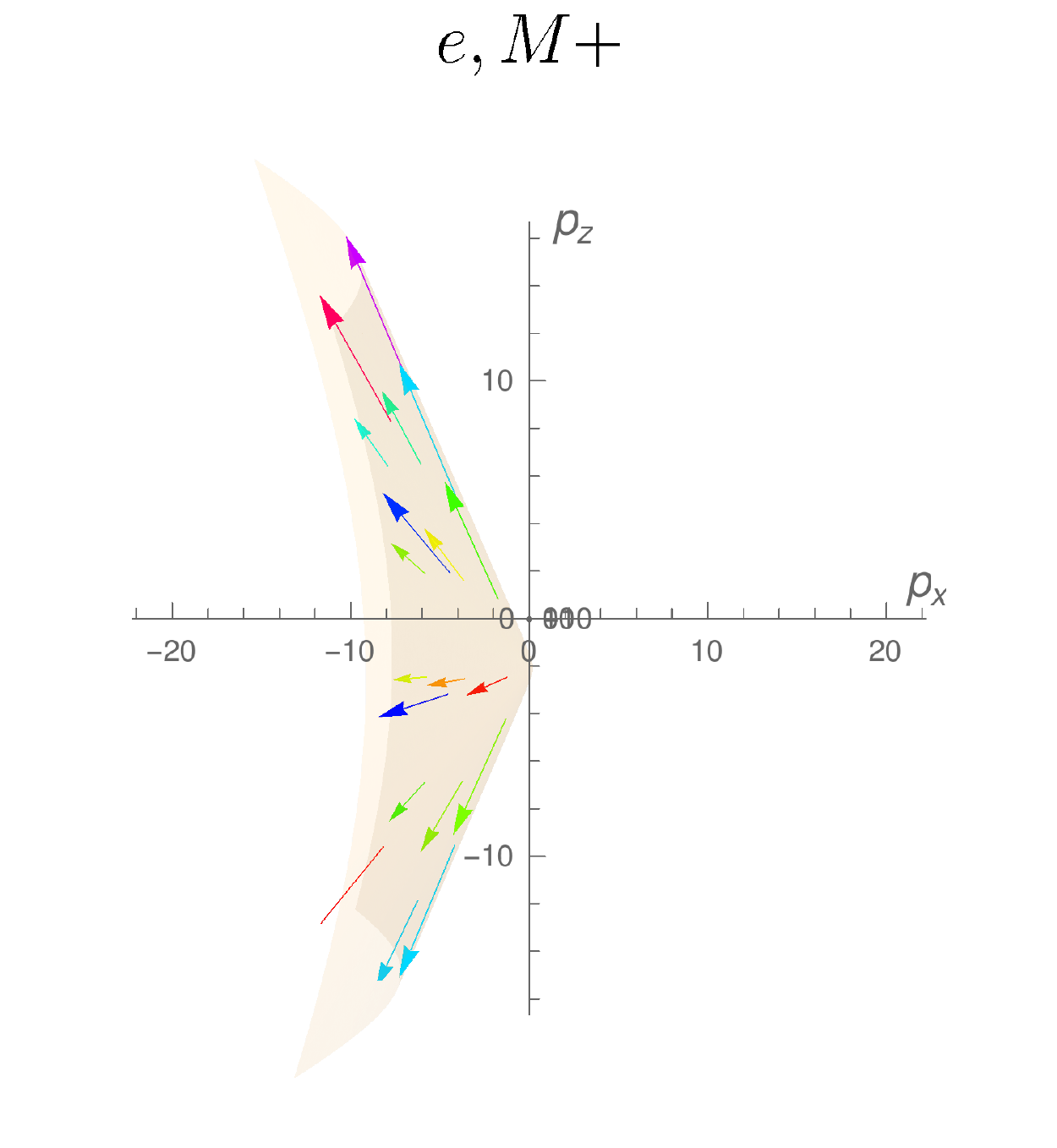}
  \hspace{.3cm}
  \includegraphics[width=0.227\textwidth]{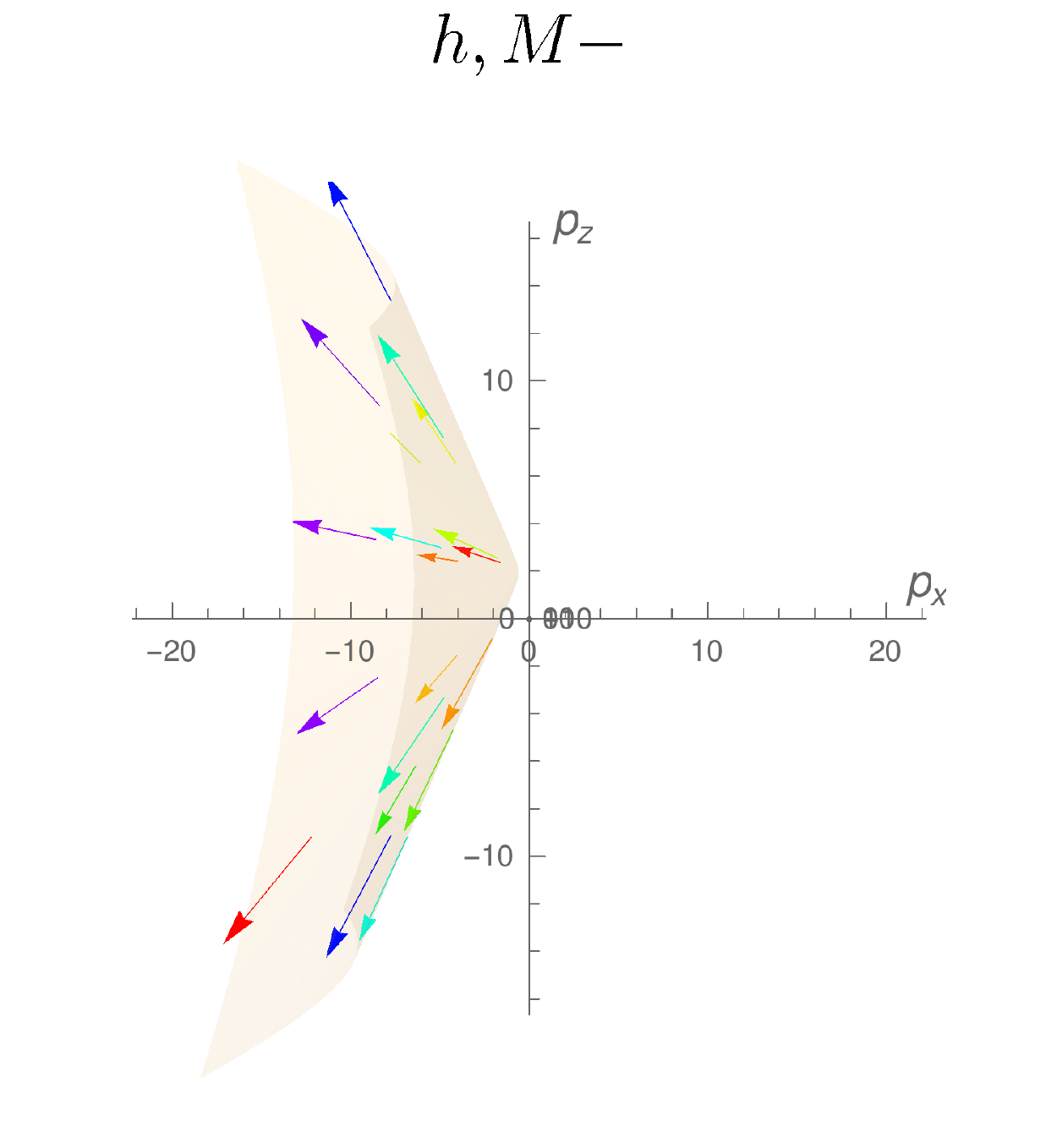}
 \hspace{.3cm}
 \includegraphics[width=0.227\textwidth]{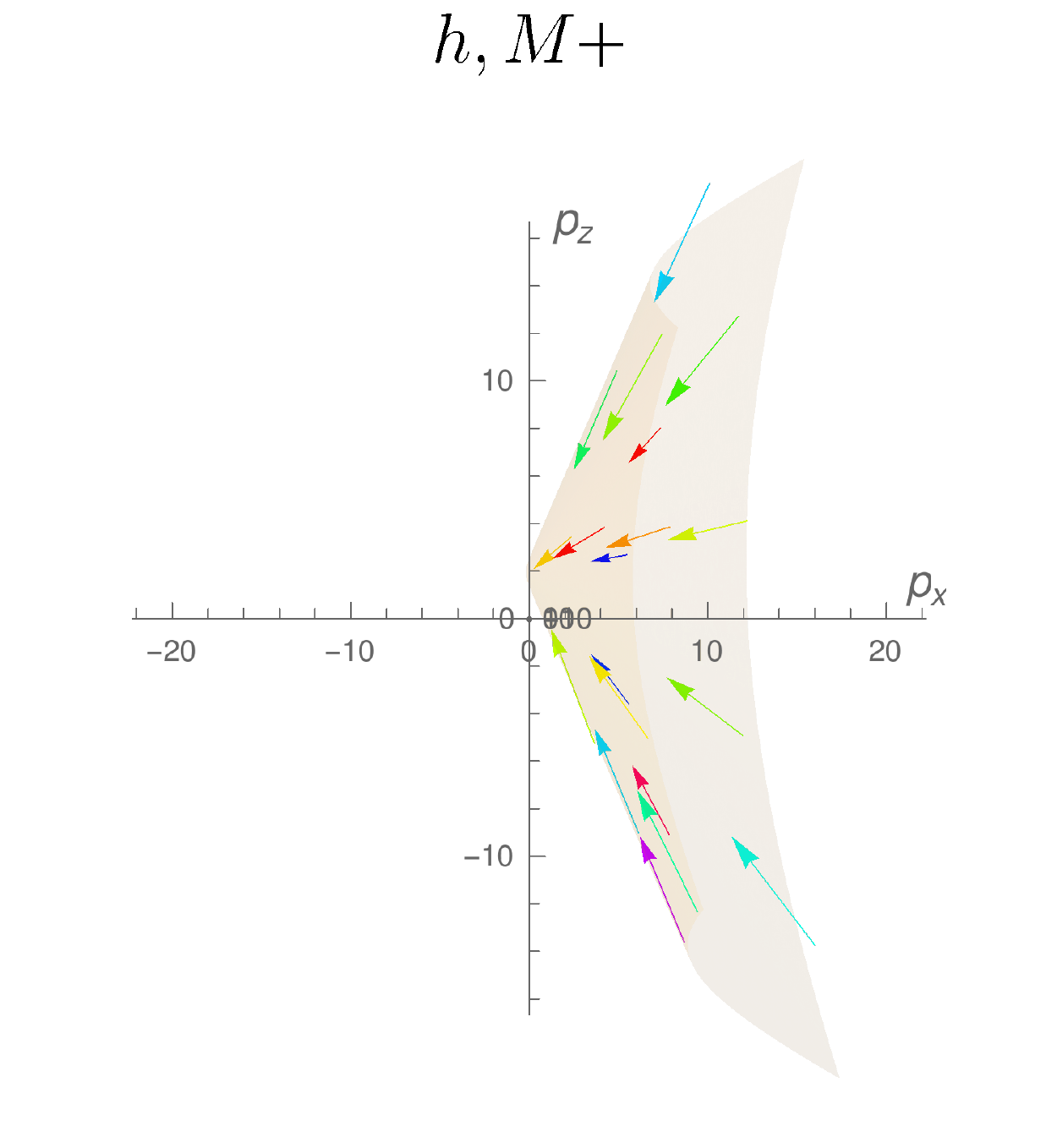}
 \caption{Plots of the average spin configuration of the incoming electrons (e$ -$) and (e$+$) and reflected holes  (h$ +$) and (h$-$)
 at the normal-magnetic interface (top row), and magnetic-superconductor interface (second and third row). The arrows points to the orientation
 of the net spin on the constant energy hyperboloid surface for a fixed bias.
 $v_1=2$, $v_2=1$, and $\mu=0.1 \Delta_0$.
 The representative magnetization along $\hat{y}$ and $\hat{z}$ direction is 0.5 and 2 for the second and third row respectively.
 The qualitative results of the spin matching in the normal and magnetic region for different branches of electron and hole
 is robust for a wide range of magnetization.}
 \label{fig4}
 \end{figure*}
 
The plots of the spin expectation values for different particle branches for a
fixed bias voltage on constant energy hyperboloid surfaces both in absence and presence of magnetic
field are given in Fig.\ref{fig4}. The analysis shows that although a $s$-wave spin singlet superconductor has been
used to induce superconducting pairing amplitudes in the junction by proximity effect, \textit {remarkably both spin singlet and spin triplet pairing
are induced in this tilted superconducting Weyl junction}. Similar observations has been made for the topological surface states, where the $s$-wave proximity
effect has been found to always induce two kind of spin pairing i.e singlet and triplet \cite{ABS2013, KZhang2017}. 
For Lorentz symmetry breaking Weyl semimetal junctions this has not been reported so far. When superconductivity is
induced in these inversion symmetry breaking material by proximity effect of conventional $s$-wave superconductor, 
the resulting superconducting ground state supports both the spin-singlet and triplet pairing. 
For the Weyl-II NMS junction, the intraband or retro Andreev reflections is marked by
spin triplet pairing while the interband or specular Andreev reflections are spin singlet pairings.
The orientation of the spin of the incoming electrons as well as the reflected holes is found to be insensitive to the
magnetization amplitude of the magnetic strips, and angle of incidence. In both normal and the magnetic region,
the spin orientation $(<S_x>, <S_y>, <S_z>)$ of the incident electrons
and reflected holes are same apart from the change in magnitude in the magnetic region.
\textit{Thus with prefect spin matching, the magnetic barrier becomes transparent to the incident electrons,
and there is no suppression of the Andreev reflected holes which contributes to the differential tunnelling conductance of the junction.
With the aid of the Andreev reflections from two hole branches, the conductance thus shows robust behaviour
in the subgap bias}. This is one of the central results of the section.

In Fig.\ref{fig3}(a) the differential conductance is plotted as function of bias voltage
for different dimensionless magnetic barrier constant $\lambda_m$,
characterizing the magnetic barrier.
$\lambda_m = L/l_m$, where $l_m^{-1} = \sqrt{2}v_2/\sqrt{v_1^2-v_2^2}\prod_{j}\sqrt{\mathcal{B}_{j}}$ is the magnetic length,
$j$ denotes the $y,z$ axes, $v_1,v_2$ are the spectrum tilt and Fermi velocity
respective in the magnetic region and $L$ is the barrier width. $\lambda_m$ can be varied in two
ways, first by fixing magnetization $\mathcal{B}_{y/z}$ (hence $l_m$) and changing the width of the barrier $L$ or
by fixing $L$ and varying the magnetization $\mathcal{B}_{y/z}$.
In Fig.\ref{fig3}(b) the differential conductance is plotted as function of bias voltage
for different chemical potential $\mu_N$ in the normal region keeping the chemical potential in the magnetic region
fixed.
The qualitative behaviour of the normalised conductance is found to be robust and independent of the
strength of the magnetic barrier and change of chemical potential in the subgap region where the normalized
conductance shows a constant value of $2$.
With the increase in the bias, the conductance decreases for both the plots.
The robust normalised subgap  conductance
survives for large barrier strength and large difference of the chemical potential in the normal
and magnetic region.

\subsubsection{Finite barrier: Normal incidence}
\label{Fb2}
The differential conductance is found to be independent of the magnetic field as well of the width of the magnetic barrier. 
Generally, it is difficult to prove this analytically owing to the complexity of the system of equations involved with the junction. 
However, we can prove the independence of the conductance for the case of normal incidence. 
The Andreev reflection coefficients $A^{\pm}_{1(2)}$ corresponding to the different incoming modes 
satisfy the following relation:
\begin{align}\label{NReq1}
A_1^{(s)}+A_2^{(s)} &=  f(E, \Delta_0), \\ 
f(E, \Delta_0) &=1 \cdot \Theta(\Delta_0-E)+\frac{1}{\Delta_0^2}\left(E-\sqrt{E^2-\Delta_0^2}\right)^2\Theta(E-\Delta_0), 
\end{align}
for incident modes $\psi^{s=\pm}_e$ and $\Theta(x)$ is the Heaviside theta function. The form of $f(E, \Delta_0)$ 
follows from the definition of $\beta$ for $E \lessgtr \Delta_0$. The summation in Eqn.(\ref{NReq1}) indicates that the Andreev reflection 
coefficients and hence the differential conductance shows a robust independence of the strength of induced magnetization, 
magnetic barrier width which has been depicted in the plots obtained in Figs.(\ref{fig2}) and (\ref{fig3}).
For simplicity, we restrict further calculations to magnetization induced 
along $\hat{y}$ direction and a specific mode of incidence $viz.$ $s=-$ and note that 
analogous calculations are applicable for the $s=+$ mode and for induced magnetizations along both $\hat{y}$ and $\hat{z}$ directions 
barring the fact that the calculations are more cumbersome and tedious in the latter cases.
To this end we define, 
\begin{align}
\label{NReq2}
\gamma_1 & \equiv |\langle\psi_{h+}|J_x|\psi_{h+}\rangle|=2p_+'(v_1p_+'-v_2p_{x+}'),\nonumber \\
\gamma_2 & \equiv |\langle\psi_{h-}|J_x|\psi_{h-}\rangle|=2p_-'(v_1p_-'+v_2p_{x-}'),\nonumber \\
 \alpha & \equiv |\langle\psi_{e-}|J_x|\psi_{e-}\rangle| =2p_-(v_1p_--v_2p_{x-}).
\end{align}
The reflection parameters $r_{1,2}$ defined in the Andreev reflection coefficients (both retro and specular)
in Eqn.(\ref{dr9}) can be rewritten as:  
\begin{equation}
\label{NReq3}
r_{1,2}=\frac{\det M_{1,2}}{\det M},
\end{equation}
in terms of matrix $M$ which is obtained from the boundary conditions in Eqn.(\ref{bc}) of the junction 
and matrix $M_{1(2)}$ corresponding to the system of equations for conditions of retro (specular) reflection respectively.
For a magnetic region of width $L$, the generic form of $\det M_{1(2)}$ and $\det M$ are as follows:
\begin{align}
\label{NReq4}
\det\,M_{1(2)}& =\sum_j u(w)_je^{{\rm i}q_j L} e^{{\rm i}(p_yy+p_zz)}, \\
\label{NReq4a}
|\det\,M_1|^2&=|u_1|^2+|u_2|^2+|u_3|^2+|u_4|^2 
+2\sum\limits_{m<n}{\rm Re}\,(u_m\bar{u}_n)\cos(q_m-q_n)L 
+2\sum\limits_{m<n}{\rm Im}\,(u_m\bar{u}_n)\sin(q_m-q_n)L, \\
\label{NReq4b}
|\det\,M_2|^2&=|w_1|^2+|w_2|^2+|w_3|^2+|w_4|^2
+2\sum\limits_{m<n}{\rm Re}\,(w_m\bar{w}_n)\cos(q_m-q_n)L 
+2\sum\limits_{m<n}{\rm Im}\,(w_m\bar{w}_n)\sin(q_m-q_n)L, \\
\label{NReq4c}
|\det\,M|^2 & =4|e^{4{\rm i}\beta}|\left[(p_-'p_{x+}'+p_+'p_{x-}')^2+p_y^2(p_+'+p_-')^2\right] 
\cdot\left[(p^m_-p^m_{x+}+p^m_+p^m_{x-})^2+(p_y+\mathcal{B}_y)^2(p^m_++p^m_-)^2\right]
\cdot \nonumber \\
&\left[(p'^m_-p'^m_{x+}+p'^m_+p'^m_{x-})^2+(p_y-\mathcal{B}_y)^2(p'^m_++p'^m_-)^2\right], 
\end{align}
where, $q_1=(p^m_{x+}+p'^m_{x-})$, $q_2=(p^m_{x+}+p'^m_{x+})$, $q_3=(p^m_{x-}+p'^m_{x-})$ and $q_4=(p^m_{x-}+p'^m_{x+})$, 
$u(w)'$s are complex functions of electron and hole momenta in different regions of the junction. 
The generic forms of functions $u(w)'$s and the different useful forms of functions of 
momenta are detailed in Appendix (\ref{A2}). Then for our purpose, it is sufficient to prove that:
\begin{align}
\label{NReq5}
\gamma_1 |\det M_1|^2 +\gamma_2 |\det M_2|^2 = \alpha f(E, \Delta_0)  |\det M|^2
\end{align}
We now impose the restriction for normal incidence i.e $p_y=p_z=0$. In this limit, 
\begin{align}\label{NReq8}
p_{\pm} &=|p_{x\pm}|, \quad p_{\pm}'=|p_{x\pm}'|, \quad p^m_{\pm}=\sqrt{(p^{m}_{x\pm})^2+\mathcal{B}_y^2}, 
\quad p'^m_{\pm}=\sqrt{(p'^{m}_{x\pm})^2+\mathcal{B}_y^2}, \quad 
|p^m_{x\pm}|=|p'^m_{x\pm}|\Rightarrow \quad p^m_{\pm} = p'^m_{\pm}.\end{align}
Using expressions of different momenta as charted out in Appendix \ref{A2}, it can be shown for any arbitrary width 
$L$ of the magnetic barrier the summations over the cosines and sine terms 
for the l.h.s in Eqn.(\ref{NReq5}) are as follows: 
\begin{eqnarray}\label{NReq6}
\sum\limits_{m<n}\left[\gamma_1{\rm Re}\,(u_m\bar{u}_n)+ 
\gamma_2{\rm Re}\,(w_m\bar{w}_n)\right]\cos(q_m-q_n)L=0, \nonumber \\
\label{NReq7}
\sum\limits_{m<n}\left[\gamma_1{\rm Im}\,(u_m\bar{u}_n)+ 
\gamma_2{\rm Im}\,(w_m\bar{w}_n)\right]\sin(q_m-q_n)L=0. 
\end{eqnarray}
Expanding the terms of the r.h.s and also the rest of the terms of l.h.s of Eqn.(\ref{NReq5}) we have:
\begin{align}\label{NReq9}
&\alpha|\det\,M|^2=32p_{x+}'^2p_{x-}'^2p_{x-}^2(v_1-v_2)[(p^m_-p^m_{x+}+p^m_+p^m_{x-})^2+\mathcal{B}_y^2(p^m_++p^m_-)^2]^2, \\
\label{NReq10}
&\gamma_1(|u_1|^2+|u_2|^2+|u_3|^2+|u_4|^2)+\gamma_2(|w_1|^2+|w_2|^2+|w_3|^2+|w_4|^2)
=32p_{x+}'^2p_{x-}'^2p_{x-}^2(v_1F_1+v_2F_2),
\end{align}
where, 
\begin{align}\label{NReq11}
F_1 & =4(p^{m}_+)^2(p^{m}_-)^2\left[2p^m_+p^m_-(p^m_+p^m_-+p^m_{x+}p^m_{x-})+ 
\mathcal{B}_y^2\left(2p^m_+p^m_--p^m_+p^m_{x-}+p^m_{x+}p^m_-+p^m_+p^m_{x+}-p^m_-p^m_{x-}\right)\right], \nonumber \\ 
F_2 &=4(p^{m}_+)^2(p^{m}_-)^2\left[-2p^m_{x+}p^m_{x-}(p^m_+p^m_-+p^m_{x+}p^m_{x-})+ 
\mathcal{B}_y^2\left(-2p^m_{x+}p^m_{x-}-p^m_+p^m_{x-}+p^m_{x+}p^m_-+p^m_+p^m_{x+}-p^m_-p^m_{x-}\right)\right],\end{align}
Substituting different forms of momenta 
in the additional terms of Eqn.({\ref{NReq10}), we find that 
\begin{align}\label{NReq12}
(v_1+v_2)(p^{m2}_{x+}+p^{m2}_{x-}+p^m_+p^m_{x+}-p^m_-p^m_{x-}-p^m_+p^m_{x-}+p^m_-p^m_{x+}) 
+2v_2p^m_+p^m_--2v_1p^m_{x+}p^m_{x-}+2v_2\mathcal{B}_y^2=0.
\end{align}
which validates the equality of Eqn.(\ref{NReq5}). 
\textit {We thus show analytically that the differential conductance will be independent of the influence 
of magnetization and other parameters of the N-M-S juntion.
We note here that the differential conductance receives contribution from all possible angles of 
incidence on the junction interface along with the normal incidence. Thus it is not straighforward 
to observe the independence of the conductance from strength of induced magnetization, magnetic barrier width 
and other system parameters. However, for other modes of incidence, it can be shown following an analogous method 
(as for the normal incidence) but more complex and tedious algebra that summation of the Andreev reflection coefficients and hence the 
differential conductance depends only on the bias voltage and the superconducting gap which corroborates our numerical findings.} 

\subsubsection{Thin barrier limit}
\label{TB}
We now consider the limit of thin barrier for which the dimensionless barrier constant $\lambda_m = L/l_m$ is defined
such that for $L \rightarrow 0, \; l_m^{-1}\rightarrow \infty$, $\lambda_m$ is finite.
Here $L$ is the barrier width, $l_m$ is the magnetic length.
In this limit, $p^m_{x \pm} L \rightarrow \mp \lambda_m$,
$p^{' m}_{x \pm} L \rightarrow \pm \lambda_m$, $\tilde{p}_{e,y} \rightarrow \alpha \; l^{-1}_m$,
$\tilde{p}_{e,z} \rightarrow -\alpha \; l^{-1}_m$,  $\tilde{p}_{h,y} \rightarrow -\alpha \; l^{-1}_m$,
$\tilde{p}_{h,z} \rightarrow \alpha \; l^{-1}_m$, and $\alpha=v_2/\sqrt{2}\sqrt{v^2_1-v^2_2}$. The boundary conditions
subsequently modified and 
a straightforward but cumbersome algebra yields the expressions for $r_1$ and $r_2$ in this limit,
\begin{align}
  r_{1} &= \frac{e^{-i (\beta +2 \lambda_m)}}{2(\alpha^2-1)} \frac{\mathcal{N}_1}
 {\mathcal{D}_1}, \label{AR1_tb}\\
  r_{2} &= \frac{e^{-i (\beta +2 \lambda_m)}}{2(\alpha^2-1)} \frac{\mathcal{N}_2}
 {\mathcal{D}_2}, \label{AR2_tb}\\
 \mathcal{N}_1&=(p_{x -}+ i p_y)(p^{'}_{-}+ p_z)(-2 e^{2 i \lambda_m}+ \mathcal{A}_1 \alpha + \mathcal{A}_2\alpha^2+\mathcal{C}\alpha^3) 
 +(-p_{-}+p_z)(p^{'}_{x -} + i p_y) (2 e^{2 i \lambda_m}+ \mathcal{A}_1 \alpha - \mathcal{A}_2\alpha^2+\mathcal{C}\alpha^3)  \nonumber\\
&  +\left[(p_{x -}+ i p_y)(p^{'}_{x -} + i p_y)(1-\alpha^2)\right.
\left.+(-p_{-}+p_z)(p^{'}_{-}+ p_z)(1+\alpha^2)\right]\alpha\mathcal{D}, \\ 
 \mathcal{N}_2&=(p_{x -}+ i p_y)(-p^{'}_{+}+ p_z)(2 e^{2 i \lambda_m}- \mathcal{A}_1 \alpha - \mathcal{A}_2\alpha^2-\mathcal{C}\alpha^3) 
 +(-p_{-}+p_z)(p^{'}_{x +} + i p_y)
  (-2 e^{2 i \lambda_m}- \mathcal{A}_1 \alpha + \mathcal{A}_2\alpha^2-\mathcal{C}\alpha^3) \nonumber\\
&  -\left[(p_{x -}+ i p_y)(p^{'}_{x +} + i p_y)(1-\alpha^2)\right. 
\left.+(-p_{-}+p_z)(-p^{'}_{+}+ p_z)(1+\alpha^2)\right]\alpha\mathcal{D}
\end{align}
with,
\begin{align}
\mathcal{A}_1&=(1+ i)- 2 e^{2 i \lambda_m}+(1-i)e^{4 i \lambda_m}, \quad 
\mathcal{A}_2 =2(1-e^{2 i \lambda_m}+e^{4 i \lambda_m}), \nonumber \\
\mathcal{C}&=(1-i)-2e^{2 i \lambda_m}+(1+i)e^{4 i \lambda_m}, \quad 
\mathcal{D}= (1+i)-2i e^{2 i \lambda_m}-(1-i)e^{4 i\lambda_m},\nonumber\\
\mathcal{D}_1&=\mathcal{D}_2=(p^{'}_{+}-p_z)(p^{'}_{x -} + i p_y)+(p^{'}_{x +} + i p_y)(p^{'}_{-}+p_z).
\end{align}

For large magnetization and small barrier width, incoming particles which are incident normally at
the interface becomes predominant contributing to the transport
properties of the junction. The tilt velocity for these particles $v_1 \simeq v_2$, for which $\alpha \rightarrow 0$.
Thus the expressions of the reflection co-efficients in Eqns.(\ref{AR1_tb}), (\ref{AR2_tb}) becomes identical to those
in the absence of any scalar or magnetic barrier and reiterates the numerical results
obtained for large magnetization and smaller barrier width which represents the thin barrier
limit.
\textit{In this limit, we show analytically that the retro and specular Andreev reflection coefficients are both independent of
magnetic field effect which is also reflected in the conductance plots.}

\section{Discussions}
\label{Disc}
For the NMS junction the subgap differential conductance is found to be robust with respect to strength
and orientation of magnetizations of the magnetic strips placed along the interfaces with the superconducting
region. Both the retro and specular Andreev reflection coefficients are supported in the magnetic region. 
Analysis of the spin-expectation values reveals that both spin singlet and triplet pairing is induced
by proximity effect with a $s$-wave superconductor. Such observations has been made for topological insulators; however,
for the tilted Weyl semimetals this is a first reporting of the co-existence of such spin pairings.

We now discuss possible experimental realizations of our proposed heterostructure of Weyl type II semimetals.
For $\rm{WTe_2}$ superconductivity can be induced on layers of thickness
varying from 20 nm- 50 nm with the use of proximity effect in a $\rm{NbSe_2/WTe_2}$ hybrid structure.
In such structures, $\Delta_0$ varies from 0.38 meV to 0.07meV \cite{Huang2018} depending on the thickness of
 $\rm{WTe_2}$, $\rm{T_c}\simeq 6.2-6.4K$, $\xi \simeq$ 30 nm. Use of
standard ferromagnetic strips can induce the desired magnetization. The strength of induced magnetization 
can be tuned by choosing films of varying thickness and width of the magnetic region can varied by choosing 
appropriate film width. 
Junctions whose lengths are of few $\mu$m can be developed and bias voltage can be tuned in order of $\simeq$ few $\mu$V. 
Variation in $\lambda_m$ for NMS junction can be achieved by keeping a fixed width L and tuning the strength of magnetization. Varying the thickness
of WSM-II layers in the superconducting region, $\xi$ can be varied in order to achieve different values of $\rm{L_s}/\xi$
for fixed $\rm{L_s}$.
However, in Weyl II junctions, the strength of induced magnetization is restricted to low values $\simeq0.15 T$
as large magnetization destroys the superconductivity in this material within the available setups \cite{Huang2018}.
However, insulating ferromagnet $EuS$ is a tentative candidate to induce high magnetic fields of the order of few teslas
in these junctions \cite{Moodera2017}. The higher strength of magnetization is particularly interesting
due to its possible interplay with spin-singlet and spin-triplet pairings, both of which occurs in these junctions
as observed. This may give to rise to unexplored magnetotransport features.

In conclusion, we have studied the differential tunnelling conductance in normal-magnetic-superconducting heterostructures of Weyl II
semimetal in the presence of magnetization modulated by magnetic strips. We showed that the differential conductance of a normal-magnetic-superconductor junction is 
robust and independent of the strength and orientation of the induced magnetization. This is in stark contrast to 
conductance properties of conventional ferromagnetic-superconductor 
junctions \cite{Beenakker1995} and also the junctions hosting Dirac fermions \cite{Mondal2010}.
Retro and specular Andreev reflected holes are found to exhibit different spin pairings. 
We have provided analytical results in the thin barrier limit and also for the normal incidence 
in case of finite magnetic barrier of NMS junctions which corroborates the numerical results. 
We have discussed possible experimental parameters relevant for these junction realizations to test our theoretical results.

\appendix
\section{Generic forms for $u$ and $w$'s}
\label{A1}
\begin{align}
u_1 &= \mathcal{T}_1 \mathcal{Q}_1 \mathcal{F}_1, \quad
u_2 = \mathcal{T}_2 \mathcal{Q}_1 \mathcal{F}_2, \quad
u_3 = \mathcal{T}_1 \mathcal{Q}_2 \mathcal{F}_3, \quad
u_4 = \mathcal{T}_2 \mathcal{Q}_2 \mathcal{F}_4, \\
w_1 &= \mathcal{T}^{'}_1 \mathcal{Q}_1 \mathcal{F}_1, \quad
w_2 = \mathcal{T}^{'}_2 \mathcal{Q}_1 \mathcal{F}_2, \quad
w_3 = \mathcal{T}^{'}_1 \mathcal{Q}_2 \mathcal{F}_3, \quad
w_4 = \mathcal{T}^{'}_2 \mathcal{Q}_2 \mathcal{F}_4, 
\end{align}

where, 
\begin{align}
 \mathcal{T}_1&=-(p^{\prime}_- p^{\prime m}_{x +}+p^{\prime}_{x -} p^{\prime m}_{+}), \quad
 \mathcal{T}_2=-(p^{\prime}_- p^{\prime m}_{x -}-p^{\prime}_{x -} p^{\prime m}_{-}), \nonumber \\
 \mathcal{Q}_1& =p_{-}p^{m}_{x-}-p_{x-}p^{m}_-, \quad \quad \;\;
 \mathcal{Q}_2=p_{-}p^{m}_{x+}+p_{x-}p^{m}_+,
\end{align}

\begin{align}
 \mathcal{F}_1&=2 e^{i \beta}(p^m_{+}p^{\prime m}_{x -} - p^{\prime m}_{-}p^{m}_{x +} +i \tilde{p}_y(p^m_+ - p^{\prime m}_{-})), \quad
 \mathcal{F}_2=2 e^{i \beta}(p^m_{+}p^{\prime m}_{x +} + p^{\prime m}_{+}p^{m}_{x +} +i \tilde{p}_y(p^m_+ + p^{\prime m}_{+})), \nonumber \\ 
 \mathcal{F}_3&=-2 e^{i \beta}(p^m_{-}p^{\prime m}_{x -} + p^{\prime m}_{-}p^{m}_{x -} +i \tilde{p}_y(p^m_- + p^{\prime m}_{-})), \;
 \mathcal{F}_4=2 e^{i \beta}(p^{\prime m}_{+}p^{m}_{x -} - p^{m}_{-}p^{\prime m}_{x +} +i \tilde{p}_y(p^{\prime m}_+ - p^{m}_{-})).
\end{align}

The functions $\mathcal{T}^{'}_j$'s are obtained by changing $p^{\prime}_- \rightarrow p^{\prime}_+$, $p^{\prime}_{x -} \rightarrow - p^{\prime}_{x +}$ in the 
expressions for $\mathcal{T}_j$'s and multiplying the expressions by an overall $''-''$ sign.

\section{Useful expressions for momenta}
\label{A2}

\begin{equation}\label{Q10}
(p^{m}_{x+})^2+(p^{m}_{x-})^2=\frac{2}{v_1^2-v_2^2}\left[\frac{v_1^2+v_2^2}{v_1^2-v_2^2}E^2+M_y^2\right],\hspace{1cm}
p^m_{x+}p^m_{x-}=\frac{E^2-M_y^2}{v_1^2-v_2^2},\end{equation}
\begin{equation}
p^m_+p^m_-
=\sqrt{((p^{m}_{x+})^2+\mathcal{B}_y^2)((p^{m}_{x-})^2+\mathcal{B}_y^2)}=\frac{E^2+\frac{v_1^2}{v_2^2}M_y^2}{v_1^2-v_2^2},\end{equation}
\begin{equation}
p^m_+p^m_{x-}-p^m_-p^m_{x+}
=\sqrt{2p^m_{x+}p^m_{x-}(p^m_{x+}p^m_{x-}-p^m_+p^m_-)+\mathcal{B}_y^2(p^{m2}_{x+}+p^{m2}_{x-})}=\frac{2v_1M_y^2}{v_2(v_1^2-v_2^2)},
\end{equation}
\begin{equation}
(p^{m}_{x+})^4+(p^{m}_{x-})^4=\frac{2}{(v_1^2-v_2^2)^4}\left[E^4(v_1^4+6v_1^2v_2^2+v_2^4)+(6v_1^2+2v_2^2)(v_1^2-v_2^2)E^2M_y^2+(v_1^2-v_2^2)^2M_y^4\right],\end{equation}
\[p^m_+p^m_{x+}-p^m_-p^m_{x-}=-\sqrt{(p^m_+p^m_{x+}-p^m_-p^m_{x-})^2}=-\sqrt{p^{m4}_{x+}+p^{m4}_{x-}+\mathcal{B}_y^2(p^{m2}_{x+}+p^{m2}_{x-})-2p^m_+p^m_-p^m_{x+}p^m_{x-}}=\]
\begin{equation}\label{Q11}=-\frac{2v_1}{(v_1^2-v_2^2)v_2}\left(\frac{2E^2v_2^2}{v_1^2-v_2^2}+M_y^2\right).\end{equation}

\bibliographystyle{apsrev4-1}
\bibliography{WSM-II_Magnetic-JoP.bib}

\end{document}